\documentclass[12pt]{article}
\usepackage{graphicx}
\usepackage{enumitem}
\usepackage{natbib} 
\usepackage{url} 
\usepackage{amsmath} 
\usepackage{amsfonts}
\usepackage{amssymb} 
\usepackage{lineno}
\usepackage{setspace}
\usepackage{lscape} 
\usepackage{float} 
\usepackage{breakcites}
\usepackage{pdfpages}
\usepackage{xcolor}
\usepackage{hyperref}
\usepackage{setspace}
\usepackage{graphicx}
\usepackage{natbib}
\usepackage{url}
\usepackage{amsmath}
\usepackage{amsfonts}
\usepackage{amssymb}
\usepackage{lineno}
\usepackage{setspace}
\usepackage{lscape} 
\usepackage{float} 
\usepackage{breakcites}
\usepackage{pdfpages}
\usepackage{bm}
\usepackage{xcolor}
\usepackage{hyperref}
\usepackage{setspace}
\usepackage{amsmath}
\usepackage{amssymb}
\usepackage{graphicx}
\usepackage{mathtools}
\usepackage{float}
\usepackage{tabularx}
\usepackage{array}
\usepackage{wrapfig}
\usepackage{multirow}
\usepackage{tabu}
\usepackage{booktabs}
\usepackage{pgfplots}
\usepackage{pgfplotstable}
\usepackage{qtree}
\usepackage{xcolor}
\usepackage{listings}
\usepackage{caption}
\usepackage{subcaption}
\usepackage{siunitx}
\usepackage{makecell}
\usepackage{bm}
\usepackage{setspace}
\usepackage{titlesec}
\usepackage{url}
\usepackage{colortbl}
\usepackage{dsfont}
\usepackage{amsthm}

\graphicspath{{Figures/}}
\newcommand{\blind}{1}

\usepackage{algorithm,algpseudocode}
\usepackage{float}

\begin{document}

\def\spacingset#1{\renewcommand{\baselinestretch}%
{#1}\small\normalsize} \spacingset{1}

\newtheorem{proposition}{Proposition}

\title{\bf A Dimension-Reduced Multivariate Spatial Model for Extreme Events: Balancing Flexibility and Scalability}
\author{Remy MacDonald$^1$, Benjamin Seiyon Lee$^1$, John Foley$^2$, and Justin Lee$^2$\hspace{.2cm}\\
$^1$Department of Statistics, George Mason University, Fairfax, VA\\
$^2$Metron, Inc., Reston, VA\\}
\date{}
\maketitle


\maketitle

\begin{abstract}

Modeling extreme precipitation and temperature is vital for understanding the impacts of climate change, as hazards like intense rainfall and record-breaking temperatures can result in severe consequences, including floods, droughts, and wildfires. Gaining insight into the spatial variation and interactions between these extremes is critical for effective risk management, early warning systems, and informed policy-making. However, challenges such as the rarity of extreme events, spatial dependencies, and complex cross-variable interactions hinder accurate modeling. We introduce a novel framework for modeling spatial extremes, building upon spatial generalized extreme value (GEV) models. Our approach incorporates a dimension-reduced latent spatial process to improve scalability and flexibility, particularly in capturing asymmetry in cross-covariance structures. This Joint Latent Spatial GEV model (JLS-GEV) overcomes key limitations of existing methods by providing a more flexible framework for inter-variable dependencies.  
In addition to addressing event rarity, spatial dependence and cross-variable interactions, JLS-GEV supports nonstationary spatial behaviors and independently collected data sources, while maintaining practical fitting times through dimension reduction.  We validate JLS-GEV through extensive simulation studies, demonstrating its superior performance in capturing spatial extremes compared to baseline modeling approaches. Application to real-world data on extreme precipitation and temperature in the southeastern United States highlights its practical utility. While primarily motivated by environmental challenges, this framework is broadly applicable to interdisciplinary studies of spatial extremes in interdependent natural processes.

\end{abstract}

\noindent%
\vfill

\newpage
\spacingset{2}

\setlength{\belowdisplayskip}{3pt} \setlength{\belowdisplayshortskip}{3pt}
\setlength{\abovedisplayskip}{3pt} \setlength{\abovedisplayshortskip}{3pt}

\section{Introduction}
\label{intro}

Modeling extreme precipitation and temperature is crucial for understanding climate change impacts \citep{dore2005climate,trenberth2011changes}, as hazards like intense rainfall and record-breaking temperatures can cause damage to property, infrastructure, and ecosystems through floods, droughts, wildfires, and other disruptions \citep{easterling2016detection,change2013physical}. Studying these events helps improve early warning systems \citep{stott2004human}, optimize resource management \citep{kundzewicz2008implications}, and inform policies to protect communities and environments \citep{seneviratne2014no}. 
Climate risk management strategies can be bolstered by understanding the spatial dependence of extreme events and the interactions between distinct environmental processes \citep{schmidt2003bayesian}.

The need for accurate, multivariate models of geospatial extremes goes beyond the environmental sciences and 
 motivates this paper's contribution (i.e., balancing modeling flexibility/complexity with scalability). Applications of these models in the agricultural, biological, and environmental sciences include analyses of heavy rainfall to support infrastructure maintenance \citep{vandeskog2022modelling}, joint spatial models of humidity and temperature to assess agricultural productivity \citep{mckinnon2020estimating}, environmental contours to identify critical pollutant combinations \citep{simpson2024inference}, and modeling cross-covariances structure between different types of soil nutrients in forest ecology \citep{guhaniyogi2013modeling}.  
Addressing the challenges of extreme events, spatial dependence, and cross-variable interactions requires expanding model complexity. However, this drive for greater modeling flexibility must be reined in by realistic (computationally scalable) approaches to maintain practical utility. After briefly reviewing these challenges, we summarize our contribution of a new multivariate spatial model for extreme events that balances flexibility and scalability.

While extreme value models focus on tail events and quantifying uncertainty for high-impact risks, univariate models often neglect spatial dependencies present in the data. To address this, spatial extreme value models, including Bayesian hierarchical models \citep{cooley2007bayesian, cooley2019univariate}, copulas \citep{demarta2005t, gudendorf2010extreme, opitz2013extremal, ribatet2013extreme}, and max-stable random fields \citep{smith1990max, davis1984tail, kabluchko2009stationary}, have been developed. Among these, spatial GEV models offer a balance between flexibility, interpretability, and uncertainty quantification.

Advances in multivariate spatial models have resulted in improved predictions, specifically by modeling relationships between variables across locations (e.g., using precipitation in one region to predict temperature in a neighboring region). Defining valid cross-covariance functions to capture such interactions remains challenging. Existing methods, including separable models \citep{brown1994multivariate,sun1998assessment}, kernel convolution \citep{higdon1999non,ver1998constructing}, and coregionalization \citep{wackernagel2003multivariate,banerjee2003hierarchical}, often impose restrictive assumptions or become computationally inefficient for large datasets. Cross-covariance functions are particularly complex in environmental processes, where asymmetry—spatial delays in one variable influencing another—is common. While some approaches address asymmetry, they often assume stationarity \citep{li2011approach,qadir2021flexible}, limiting flexibility.

We develop a new approach for modeling multivariate spatial extreme data that scales to large datasets and offers additional flexibility in the cross-covariance structure: the Joint Latent Spatial GEV model (JLS-GEV). JLS-GEV's flexibility to accommodate both nonstationary spatial dependence \citep{genton2015cross} and asymmetric cross-covariances as well as measurements collected at completely distinct locations contribute to its flexibility. We demonstrate the advantages of our proposed method through simulation studies and apply it to real-world data on extreme precipitation and temperature in the southeastern United States. Our proposed methods will enable practitioners to model large-scale multivariate extreme events across space, which may have been prohibitive using existing methods; thus, striking a balance between model flexibility and scalability.
%

The remainder of this manuscript is organized as follows: Section~\ref{sec:motivation} introduces the data and scientific motivation for this study. Section~\ref{sec:stat_models} reviews existing methods for modeling climate and extreme weather (CEW) hazards and multivariate spatial processes. In Section~\ref{sec:ch4_methodology}, we propose our approach and address limitations in the current literature. Section~\ref{sec:ch4_SimulationStudy} presents simulation results, and Section~\ref{sec:ch4_Applications} illustrates our method's application to real-world data. Finally, Section~\ref{sec:ch4_discussion} provides concluding remarks and directions for future research.

\section{Data and Scientific Motivation}
\label{sec:motivation}

The southeastern US, comprised of Florida, Georgia, Alabama, Mississippi, North Carolina, South Carolina, and Tennessee, has experienced a myriad of extreme weather events such as floods, droughts, heat waves, and storm surges due to tropical cyclones \citep{konrad2013climate}. From 1980, these events have caused more billion-dollar weather relief efforts in the southern region than any other region in the US \citep{konrad2013climate}. We focus on extreme precipitation and temperatures in the southeastern United States. Jointly modeling these two weather- and climate-related hazards can offer valuable insights and guide decision-makers in designing adaptive resilience strategies.


The data are obtained from the Global Historical Climatology Network Daily (GHCND) \citep{menne2012overview}, a comprehensive climate database maintained by the National Centers for Environmental Information (NCEI), part of the National Oceanic and Atmospheric Administration (NOAA). The GHCND includes records from over 100,000 weather stations across 180 countries and territories, capturing variables such as maximum and minimum temperature, daily precipitation, snowfall, and snow depth. Quality assurance checks are routinely applied, and the GHCND had been cited 1,945 times as of December 6, 2024. In the supplement, we provide details on our data acquisition process using public interfaces provided by NOAA and other open-source software.   

\begin{figure}[H]
\begin{center}
\begin{tabular}{cc}
\multicolumn{2}{c}{\includegraphics[width=100mm]{Figures/Map_Stations.pdf}}  \\
\multicolumn{2}{c}{(a) Location of precipitation stations (red) and temperature stations (black).}  \\[20pt]
\includegraphics[width=58mm]{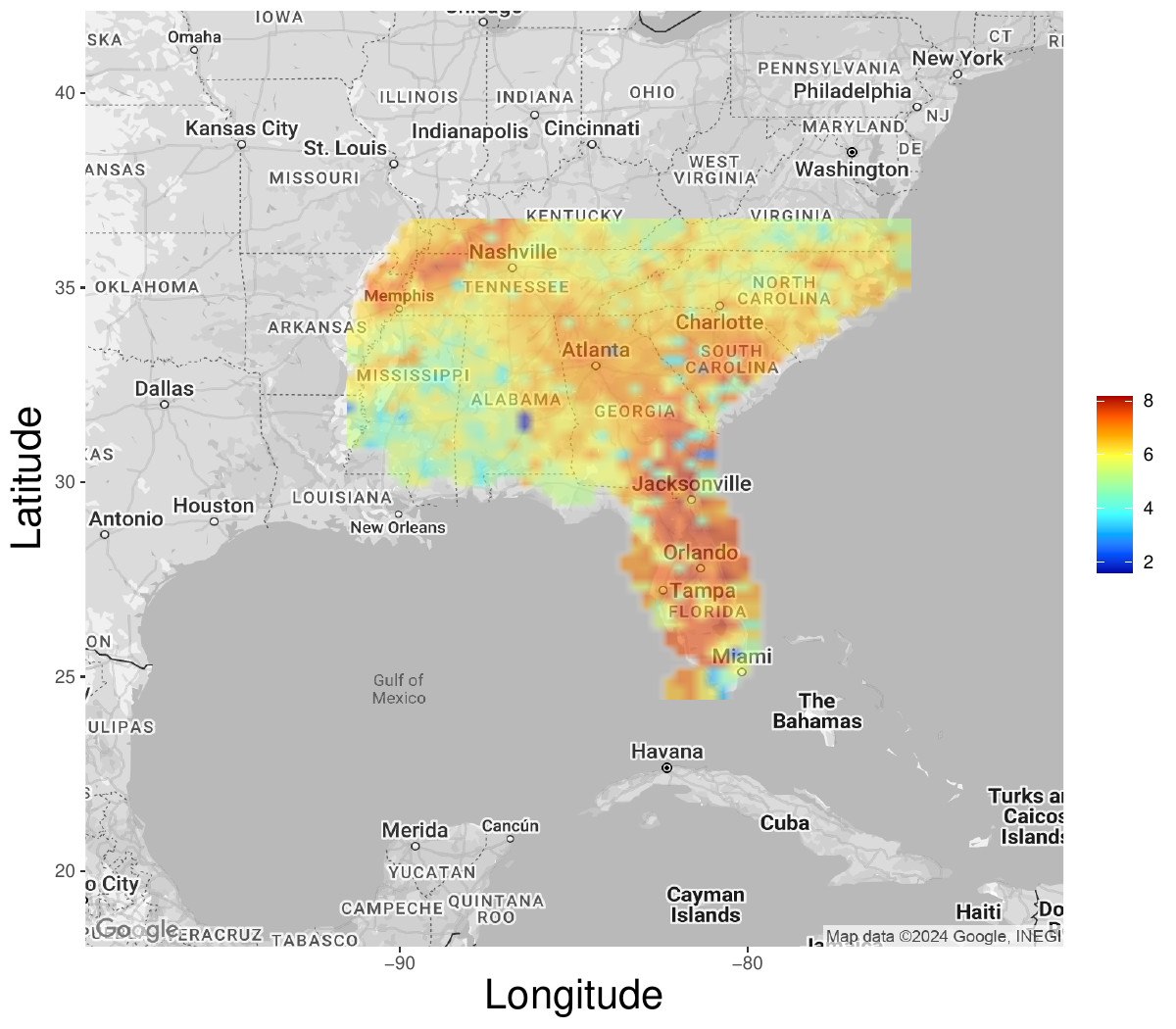} &   \includegraphics[width=58mm]{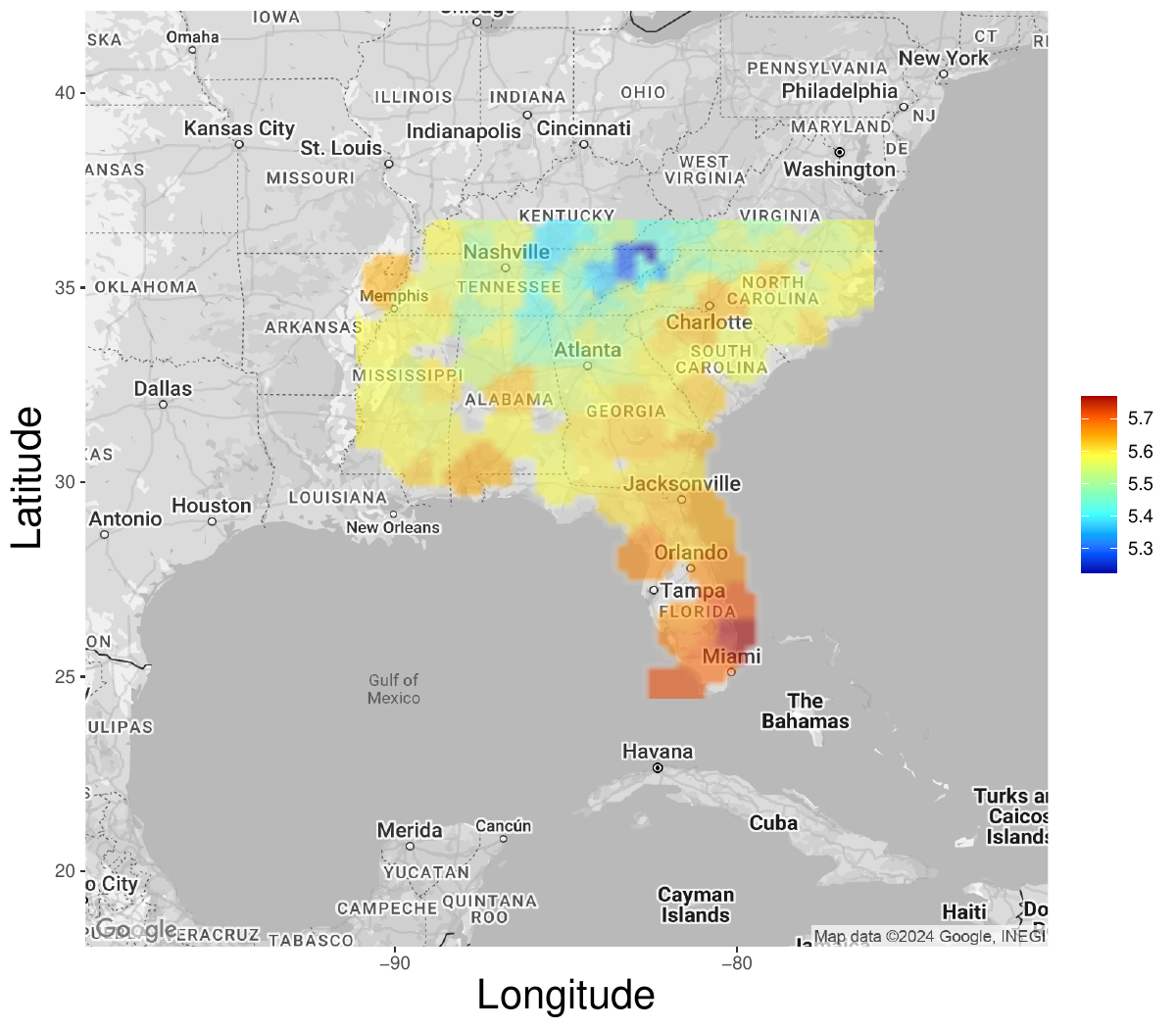} \\
(b) Log maximum precipitation & (c) Log maximum temperature \\[20pt]
\end{tabular}
\caption{(a) Locations of precipitation stations (red) and temperature stations (black). (b) Log of maximum daily precipitation in tenths of millimeters and (c) maximum average daily temperature in tenths of degrees Celsius from September 2017, which coincides with the landfall of Hurricane Irma.
}
\label{Fig3:Map_Stations}
\end{center}
\end{figure}

Daily precipitation data, in tenths of millimeters, is available for weather stations in the GHCND. There are 10,610 unique precipitation stations across the southeastern U.S. dating back to 1852. For each station, we extract the monthly maximum values as block-maxima. The locations of these weather stations are represented by the red points in Figure \ref{Fig3:Map_Stations}. Daily average temperature, in tenths of degrees Celsius, is available at 248 unique temperature stations dating back to 1892. In Figure \ref{Fig3:Map_Stations}, the temperature stations are represented as black points. Figure \ref{Fig3:Map_Stations} also displays a snapshot if the natural log of the maximum daily precipitation and temperature recorded during the landfall of Hurricane Irma (September 2017). In this real-world example, we see the need to jointly model two extreme weather hazards while address two key challenges - processing millions 
of high-quality measurements and reconciling distinct locations.

\section{Statistical Models for Extreme Weather Events}
\label{sec:stat_models}

The development of statistical models for extreme weather and climate hazards has been a dynamic area of research, particularly in storm surge modeling \citep{vousdoukas2016projections}, precipitation analysis \citep{donat2016more}, wind studies \citep{martius2016global}, and more. In the section, we review the progression of models that our work builds on, culminating in a discussion of outstanding challenges for our present study of extreme precipitation and temperature. 

Extreme weather and climate hazards hazards often exhibit characteristics like left-bounded constraints (e.g., non-negative values), right-skewed distributions with heavy tails, and heterogeneous tail behavior across regions, making traditional models inadequate. To capture heavy-tailed events, extreme value models have been developed, employing techniques like block maxima \citep{coles2001introduction} and peaks-over-threshold \citep{pickands1975statistical}. However, these univariate models assume independence among observations, neglecting spatial dependence.

To address this, various spatial extremes models have emerged, incorporating spatial autocorrelation. Copula-based models, particularly extremal $t$-copulas \citep{demarta2005t, gudendorf2010extreme, opitz2013extremal, ribatet2013extreme}, handle spatial dependence but lack the max-stable property. Spatial GEV models \citep{cooley2007bayesian, cooley2019univariate} introduce spatial dependence through varying parameters represented by Gaussian processes (GPs) or basis expansions. Max-stable models, like the Smith \citep{smith1990max} or Brown-Resnick \citep{davis1984tail, kabluchko2009stationary} models, ensure critical asymptotic properties of block maxima.

Despite their benefits, spatial extreme models face a tradeoff between complexity and practicality. Max-stable models are flexible but can be slow to fit and challenging to scale for large datasets. Both max-stable and copula-based models require extensive computation of the multivariate GEV distribution’s likelihood, which can be computationally prohibitive \citep{castruccio2016high}. Pairwise likelihood functions \citep{genton2011likelihood} replaces the original likelihood by a pseudo-likelihood constructed from pairwise likelihoods; however, it is difficult to impose an interpretable sparse structure on the composite likelihood \citep{yu2016modeling}.

\subsection{Generalized Extreme Value (GEV) Model}
\label{sec:EVT}
Extreme value theory (EVT) is a branch of statistics focused on characterizing the tail behavior of a distribution. The foundation of EVT is built on asymptotic results that yield distributions used to describe the tails. In practice, large observations, such as annual maxima (block maxima) or exceedances over a high threshold, can be used to fit these distributions. The generalized Pareto distribution (GPD) has been used to model exceedance-over-threshold data. While these methods allow the use of more data, it can be challenging to choose an appropriate cut-off and to de-cluster dependent data \citep{davison1990models}. In this study, we focus on modeling block maxima, which is amenable to the generalized extreme value (GEV) distribution \citep{coles2001introduction}. 

The GEV distribution is defined as:
\begin{equation}\label{eq:GEV}
P(X \leq x) = \exp\left\{-\left[1 + \xi\left(\frac{x - \mu}{\sigma}\right)\right]^{-1/\xi}\right\},
\end{equation}

\noindent where \(1 + \xi\left(\frac{x - \mu}{\sigma}\right) > 0\). The parameters \(\mu\) and \(\sigma\) represent the location and scale parameters, respectively, while the shape parameter \(\xi\) determines the tail behavior. If $\xi>0$ the upper tail is bounded, if $\xi=0$ then (\ref{eq:GEV}) is understood as the limit and the tail decays exponentially, and if $\xi<0$ then the tail decays as a power function. In practice, a GEV distribution is fitted to block maxima (e.g., annual maxima) extracted from a long data record. The fitted model can provide quantities of interest such as return levels and periods. 

\subsection{Spatial Generalized Extreme Value Model}
\label{sec:spatial_GEV}

Spatial GEV models are commonly used to model extreme weather events, particularly when accounting for spatial dependence across locations \citep{cooley2007bayesian,davison2012statistical}. These models extend the traditional GEV distribution by incorporating spatial correlation, allowing for joint modeling of extremes across multiple sites \citep{padoan2010likelihood}. For example, \citet{cooley2007bayesian} used spatial GEV models to generate precipitation return level maps and uncertainty estimates for Colorado, while \citet{cooley2010spatial} applied similar models to characterize extreme precipitation simulated by a regional climate model.
%

In the spatial GEV model, the parameters of the GEV distribution—location ($\mu$), scale ($\sigma$), and shape ($\xi$)— vary across space. Due to difficulties in estimating the shape parameter, $\xi$ is often held constant across locations \citep{sang2009hierarchical}. Spatial dependence is typically modeled through a latent process, such as a Gaussian process (GP), which captures the correlation structure between locations.
By assuming that observations at different locations are conditionally independent, given the spatially varying GEV parameters, each site can be modeled with its own univariate GEV distribution.
%
%
The corresponding Bayesian hierarchical model is: 

\begin{equation}\label{EQ:spatial_extreme}
\begin{aligned}
&\textbf{Data Model: } & Z(s)|\cdot \sim \text{GEV}(\mu(s), \sigma(s),\xi(s))\\
&\textbf{Process Model: } & \log\sigma(s)\sim\mathcal N(X(s)\beta_\sigma(s)+\omega_\sigma(s),\tau_\sigma^2) \\
&& \log\mu(s)\sim\mathcal N(X(s)\beta_\mu(s)+\omega_\mu(s),\tau_\mu^2) \\
&&\omega_\sigma(s)\sim\mathcal{GP}(0,K_\sigma(\theta_\sigma))\\
&&\omega_\mu(s)\sim\mathcal{GP}(0,K_\mu(\theta_\mu))\\
&\textbf{Parameter Model: } & \beta_\sigma(s),\beta_\mu(s),\tau_\sigma^2,\tau_\mu^2,\theta_\sigma,\theta_\mu,\xi,
\end{aligned}
\end{equation}

\bigskip

\noindent where the prior distributions for $\beta_\sigma(s)$, $\beta_\mu(s)$, $\tau_\sigma^2$, $\tau_\mu^2$, $\theta_\sigma$, $\theta_\mu$, and $\xi$ need to be specified by the practitioner. Based on our exploratory data analysis, we have found that the hierarchical spatial GEV model provides a nice balance between model sophistication and computational efficiency; hence, it serves as the foundation for our proposed methodology.

\subsection{Multivariate Spatial Models}
\label{sec:multivariate}



Statistical analyses of multivariate measurements has come to the forefront in spatial statistics \citep{gelfand2021multivariate, schliep2013multilevel}. These data often exhibit dependencies both within measurements at a single location and across different locations. 

The $p$-variate observational GP model under consideration is
$$\begin{pmatrix} Y_1(\boldsymbol s)\\ \vdots\\ Y_p(\boldsymbol s)\end{pmatrix}=\begin{pmatrix} \mu_1(\boldsymbol s)\\ \vdots \\ \mu_p(\boldsymbol s)\end{pmatrix}+\begin{pmatrix} w_1(\boldsymbol s)\\ \vdots \\ w_p(\boldsymbol s)\end{pmatrix}+\begin{pmatrix} \epsilon_1(\boldsymbol s)\\ \vdots \\ \epsilon_p(\boldsymbol s)\end{pmatrix}$$

\noindent\sloppy or in vector form, $\boldsymbol Y(\boldsymbol s) = \boldsymbol\mu(\boldsymbol s) + \boldsymbol w(\boldsymbol s) + \boldsymbol\epsilon(\boldsymbol s)$. Here, $\boldsymbol Y(\boldsymbol s) =
(Y_1(\boldsymbol s),\ldots,Y_p(\boldsymbol s))^\top$ is the observed process at location $\boldsymbol s\in\mathbb R^d$ on a spatial domain $\mathcal D$
with mean $\boldsymbol\mu(\boldsymbol s) = (\mu_1(\boldsymbol s),\ldots,\mu_p(\boldsymbol s))^\top$ and spatially correlated component $\boldsymbol w(\boldsymbol s) = (w_1(\boldsymbol s),\ldots, w_p(\boldsymbol s))^\top$, which we assume to be a multivariate GP. The observations are subject to noise $\boldsymbol\epsilon(\boldsymbol s) = (\epsilon_1(\boldsymbol s),\ldots,\epsilon_p(\boldsymbol s))^\top$, a mean zero multivariate white noise process with covariance matrix
$\text{Cov}(\boldsymbol\epsilon(\boldsymbol s), \boldsymbol\epsilon(\boldsymbol s)) = \text{diag}(\tau_1^2,\ldots,\tau_p^2)$. The vector-valued spatial process $\{\boldsymbol w(\boldsymbol s)\in\mathbb R^p : \boldsymbol s\in\mathcal D\}$ is specified by the matrix-valued cross-covariance function $\boldsymbol C(\boldsymbol s, \boldsymbol s')$, which determines the joint dependence structure of the spatial process.  This matrix function must be carefully constrained in order to be a nonnegative definite matrix function. In particular, for arbitrary locations $\{\boldsymbol s_i\}_{i=1}^n$, the block matrix $\boldsymbol\Sigma$ with $(i, j)$-th block $\boldsymbol C(\boldsymbol s_i,\boldsymbol s_j)$ must be nonnegative definite.

Several methods have been developed for multivariate observations \(\boldsymbol Y = (\boldsymbol Y(\boldsymbol s_1), \ldots, \boldsymbol Y(\boldsymbol s_n))^\top\). Separable spatial models \citep{brown1994multivariate,sun1998assessment}  construct cross-covariance matrices represented as \(\boldsymbol\Sigma_{\boldsymbol Y} = \boldsymbol R \otimes \boldsymbol T\), where \(\boldsymbol R\) contains spatial correlations \(\rho(\boldsymbol s_i, \boldsymbol s_j)\) and \(\boldsymbol T\) represents measurement correlations. These models ensure positive definiteness and simplify matrix operations, but they assume uniform spatial correlation structures across components of \(\boldsymbol w(\boldsymbol s)\); thus, limiting flexibility \citep{gelfand2021multivariate}. Kernel convolution methods \citep{higdon1999non,paciorek2003nonstationary} generate valid stationary cross-covariograms using kernel functions \(k_l(\cdot)\) and a zero-mean GP \(w(\boldsymbol s)\). The spatial process $Y_l(\boldsymbol s)$ is expressed as $Y_l(\boldsymbol s) = \sigma_l \int k_l(\boldsymbol s - \boldsymbol t) w(\boldsymbol t) d\boldsymbol t,$ resulting in a valid stationary cross-covariance function using Gaussian \citep{higdon1999non} or Matérn \citep{paciorek2003nonstationary} kernels. Convolutions of one-dimensional stationary covariance functions \(C_i\) and \(C_j\) \citep{majumdar2007multivariate,gaspari1999construction} can be used to construct cross-covariance structures. However, expensive Monte Carlo integration methods may be needed \citep{gelfand2021multivariate} to construct the cross-covariance $C_{ij}(\boldsymbol{s}) = \int C_i(\boldsymbol{s} - \boldsymbol{t}) C_j(\boldsymbol{t}) d\boldsymbol{t}$.

A notable case is the linear model of coregionalization (LMC) \citep{wackernagel2003multivariate,grzebyk1994multivariate,banerjee2003hierarchical,schmidt2003bayesian}, which models multivariate processes through linear combinations of independent spatial processes $\boldsymbol Y(\boldsymbol s) = \boldsymbol A \boldsymbol w(\boldsymbol s),$ where \(\boldsymbol A\) is a full-rank matrix and \(\boldsymbol w(\boldsymbol s)\) are independent and identically distributed (i.i.d.) processes. This yields a separable cross-covariance matrix, \(C(\boldsymbol s - \boldsymbol s') = \rho(\boldsymbol s - \boldsymbol s') \boldsymbol A \boldsymbol A^\top\). LMC can be generalized to the case when $\boldsymbol w(\boldsymbol s)$ consists of independent but not identically distributed processes, each with a distinct correlation function \(\rho_j(\boldsymbol h)\), resulting in $C(\boldsymbol s - \boldsymbol s') = \sum_{j=1}^p \rho_j(\boldsymbol s - \boldsymbol s') \boldsymbol T_j,$ where \(\boldsymbol T_j = \boldsymbol a_j \boldsymbol a_j^\top\). The nonstationary extension, spatially varying LMC (SVLMC) \citep{gelfand2004nonstationary}, adapts by replacing \(\boldsymbol A\) with \(\boldsymbol A(\boldsymbol s)\), leading to $C(\boldsymbol s,\boldsymbol s') = \sum_j \rho_j(\boldsymbol s - \boldsymbol s') \boldsymbol a_j(\boldsymbol s) \boldsymbol a_j^\top(\boldsymbol s')$.

\subsection{Joint Spatial GEV Models and Challenges}
\label{jointChall}

Let $\boldsymbol Z_1=(\boldsymbol Z_1{(\boldsymbol s_1)^\top},\ldots,\boldsymbol Z_1(\boldsymbol s_n)^\top)^\top$ and $\boldsymbol Z_2=(\boldsymbol Z_2{(\boldsymbol t_1)^\top},\ldots,\boldsymbol Z_2(\boldsymbol t_m)^\top)^\top$ denote the precipitation and temperature observations, respectively, at spatial locations $\boldsymbol s_1,\ldots,\boldsymbol s_n$ and $\boldsymbol t_1,\ldots,\boldsymbol t_m$. In our case, we consider these to be the block-maxima across months. Note that observations can be collected at different locations for each process. 

The spatial GEV model represents the data as
$$Z_1(\boldsymbol s)\sim\text{GEV}(\mu_1(\boldsymbol s),\sigma_1(\boldsymbol s),\xi_1)$$
$$Z_2(\boldsymbol t)\sim\text{GEV}(\mu_2(\boldsymbol t),\sigma_2(\boldsymbol t),\xi_2),$$
\noindent where $\mu_1(\cdot)$, $\mu_2(\cdot)$, $\sigma_1(\cdot)$, and $\sigma_2(\cdot)$ represent latent spatial processes for the corresponding location and scale parameters for the precipitation and 
temperature
data. These are typically modeled as GPs. The latent process for the location parameter is:
\begin{equation}\label{eq2}
\begin{pmatrix}
\boldsymbol \mu_{1}\\ \boldsymbol \mu_{2}
\end{pmatrix}\sim\mathcal N\left(\begin{pmatrix}
\boldsymbol 0\\ \boldsymbol 0
\end{pmatrix},\begin{pmatrix}
\boldsymbol \Sigma_1 & \boldsymbol \Sigma_{12}\\ \boldsymbol \Sigma_{21} & \boldsymbol \Sigma_2
\end{pmatrix}\right),
\end{equation}
\noindent where $\boldsymbol\Sigma_1$ and $\boldsymbol\Sigma_2$ are $n\times n$ and $m\times m$ covariance matrices the spatially-varying location parameters for the precipitation and temperature data, respectively. The latent process for the scale parameter can be modeled similarly with a log link function to ensure positivity.

\paragraph{Scalability} Despite the flexibility of this model, our dataset of interest consists of \(n + m = 10{,}858\) spatial locations, making it computationally prohibitive to fit the model directly. Evaluating the hierarchical model in equation (\ref{eq2}) requires \( \mathcal{O}((n + m)^3)\) operations. In the following subsection, we will propose a basis function representation approach to reduce dimensionality and improve scalability for large datasets.

\paragraph{Stationarity} 
Assuming a stationary covariance structure for the latent spatial processes may not be sensible or realistic. The SVLMC \citep{gelfand2004nonstationary} handles nonstationarity by allowing the LCM coefficients to be space-dependent. In particular, $\boldsymbol T(\boldsymbol s) = \boldsymbol A(\boldsymbol s)\boldsymbol A(\boldsymbol s)^\top$ is modeled as a spatial process with an inverse Wishart distribution. However, it is not clear if a simple approach (LCM) would perform as well as this complex model. An alternative framework could be achieved by representing the multivariate spatial process as
$\boldsymbol v(\boldsymbol s) = \boldsymbol A\boldsymbol w(\boldsymbol s)$, with $\boldsymbol w(\boldsymbol s) = (w_1(\boldsymbol s),\ldots, w_p(\boldsymbol s))$, and treating each $w_i(\boldsymbol s)$ for $i = 1,\ldots,p$, which is a nonstationary random field modeled using a basis expansions. We employ this approach to address the nonstationarity in $\boldsymbol v$; in fact, this strategy may be easier to implement in practice as opposed to allowing the components of $\boldsymbol A$ to change with space.



\paragraph{Asymmetry} 
Existing approaches assume symmetric relationships between environmental processes, where $\mathsf{Cov}(Z_i(\boldsymbol s), Z_j(\boldsymbol s')) = \mathsf{Cov}(Z_i(\boldsymbol s'), Z_j(\boldsymbol s)))$ for processes $i$ and $j$ at locations $\boldsymbol s$ and $\boldsymbol s'$. This implies a change in the value of cross-covariance when the locations are interchanged for a fixed order of variables \citep{qadir2021flexible}. Asymmetric relationships occur due to spatial delays in one variable influencing another, which commonly occurs in environmental processes. Neglecting assymetry can lead to misrepresentation of the underlying spatial processes. \citet{apanasovich2010cross} develops latent dimension-based multivariate spatio-temporal cross-covariance models that incorporate asymmetry by adjusting latent dimensions for each variable. The conditional approach introduced by \citet{cressie2016multivariate} specifies asymmetric cross-covariances using interaction functions. \citet{li2011approach} introduce asymmetry in any valid stationary cross-covariance function by defining a new covariance function \(\boldsymbol{C}^{LZ}(\boldsymbol{h})\) that incorporates delays via vectors \(\boldsymbol{l}_i\), allowing the cross-covariances to occur at non-zero lags. 



\section{Our Approach}
\label{sec:ch4_methodology}

To address the computational and modeling challenges associated with multivariate spatial extreme models, we develop a new approach that: (i) scales to large, high-dimensional datasets; (ii) offers additional flexibility in the cross-covariance structure; and (iii) accommodates each process having distinct spatial locations. 

To reduce computation time, we employ a semi-parametric approach to approximate the latent spatial surfaces of the spatially-varying location and scale parameters via basis expansions. For this implementation, we utilize multi-resolution bisquare basis functions \citep{sengupta2016predictive} and eigenfunctions \citep{holland1999spatial}. 
%
We induce dependence across basis coefficients from each process, resulting in an asymmetric cross-covariance without stationary assumptions. Based on our study results, this specification yields a positive semidefinite covariance structure for the joint latent spatial processes. 

To accommodate data with out-of-sync measurement locations, our approach allows distinct spatial locations for each process. Many existing methods assume fully observed outputs at each spatial location. For example, \citet{lee2023class} model the cross-covariance between occurrence and prevalence processes in the context of zero-inflated spatial modeling but require the same locations for each process to construct a positive semidefinite covariance structure. Our approach overcomes this limitation while controlling model dimensionality and maintaining a positive semidefinite covariance
structure. 



\subsection{ Modeling Framework: Joint Latent Spatial GEV (JLS-GEV)}
\label{framework}

Let $\boldsymbol Z=(\boldsymbol Z_1^\top,\boldsymbol Z_2^\top)^\top$ denote the combined vectors of observations for the precipitation and 
temperature
data. Let $(\boldsymbol s,\boldsymbol t)$ denote the combined vector of spatial locations from the precipitation and 
temperature
processes, where $\boldsymbol s=(\boldsymbol s_1,\ldots,\boldsymbol s_n)$ and $\boldsymbol t=(\boldsymbol t_1,\ldots,\boldsymbol t_m)$. For brevity, let $r=n+m$ denote the total number of spatial locations. We model the data as:
$$\boldsymbol Z(\boldsymbol s) \sim \text{GEV}(\boldsymbol\mu(\boldsymbol s), \boldsymbol\sigma(\boldsymbol s), \boldsymbol\xi),$$
\noindent where \(\boldsymbol Z(\boldsymbol s) = (Z_1(\boldsymbol s), Z_2(\boldsymbol s))\) represent the extreme precipitation and temperature random variables at location \(\boldsymbol s\), with spatially varying location and scale parameters $\boldsymbol\mu(\boldsymbol s) = (\mu_1(\boldsymbol s), \mu_2(\boldsymbol s))^\top$, $\boldsymbol\sigma(\boldsymbol s) = (\sigma_1(\boldsymbol s), \sigma_2(\boldsymbol s))^\top$, and shape parameters $\boldsymbol\xi = (\xi_1, \xi_2)^\top$. 


The location and scale parameters \(\boldsymbol\mu(\boldsymbol s)\) and \(\boldsymbol{\sigma}(\boldsymbol{s})\) are modeled as:
\[
\boldsymbol\mu(\boldsymbol s) = \boldsymbol\mu_0 + \boldsymbol A_{\mu} \begin{pmatrix}
\boldsymbol\Phi_{\mu_1}^\top(\boldsymbol s)\boldsymbol\delta_{\mu_1}\\ 
\boldsymbol\Phi_{\mu_2}^\top(\boldsymbol s)\boldsymbol\delta_{\mu_2}
\end{pmatrix},
\]

\[
\boldsymbol{\sigma}(\boldsymbol{s}) = \exp\left(\boldsymbol{\sigma}_0 + \boldsymbol{A}_{\sigma} \begin{pmatrix}
\boldsymbol{\Phi}_{\sigma_1}^\top(\boldsymbol{s}) \boldsymbol{\delta}_{\sigma_1} \\
\boldsymbol{\Phi}_{\sigma_2}^\top(\boldsymbol{s}) \boldsymbol{\delta}_{\sigma_2}
\end{pmatrix}\right),
\]

\noindent where \(\boldsymbol A_{\mu}\) is a $2\times 2$ lower triangular matrix with LMC coefficients $a_{11,\mu}$, $a_{21,\mu}$, and $a_{22,\mu}$, and \(\boldsymbol A_{\sigma}\) is defined similarly with distinct LMC coefficients. For observed spatial locations \((\boldsymbol s,\boldsymbol t)\), the latent spatial processes are represented as:
\[
\boldsymbol \mu = \boldsymbol\mu_{0} + \boldsymbol A_{\mu}^* \boldsymbol\Phi_{\mu} \boldsymbol\delta_{\mu},
\]
\[
\boldsymbol \sigma = \exp(\boldsymbol\sigma_{0} + \boldsymbol A_{\sigma}^*\boldsymbol\Phi_{\sigma} \boldsymbol\delta_{\sigma}),
\]
\noindent where \(\boldsymbol\mu_{0}\) and \(\boldsymbol\sigma_{0}\) are offset parameters. The $r\times 2r$ block matrix $\boldsymbol A_{\mu}^*$ is defined as:
$$\boldsymbol A_{\mu}^* = \begin{pmatrix}
\boldsymbol A_{11,\mu} & \boldsymbol 0 \\
\boldsymbol A_{21,\mu} & \boldsymbol A_{22,\mu} \\
\end{pmatrix},$$

\noindent with \(\boldsymbol A_{11,\mu}\), \(\boldsymbol A_{21,\mu}\), and \(\boldsymbol A_{22,\mu}\) structured to incorporate the respective LMC coefficients for the location process. The $r\times 2r$ block matrix $\boldsymbol A_\sigma^*$ is defined similarly. Additionally, we define $\boldsymbol\Phi_\mu$ to be a $2r\times(p+q)$ basis function matrix,
$$\boldsymbol\Phi_\mu=\begin{pmatrix}
    \boldsymbol\Phi_{\mu_1} & \boldsymbol 0 \\
    \boldsymbol0 & \boldsymbol\Phi_{\mu_2}
\end{pmatrix},$$

\noindent where $\boldsymbol\Phi_{\mu_1}$ is an $r\times p$ basis function matrix with corresponding $p$-vector of basis coefficients $\boldsymbol\delta_{\mu_1}$, and $\boldsymbol\Phi_{\mu_2}$ is an $r\times q$ basis function matrix with corresponding $q$-vector of basis coefficients $\boldsymbol\delta_{\mu_2}$. For the combined basis function matrix $\boldsymbol\Phi_\mu$, the corresponding $(p+q)$-vector of basis coefficients is denoted by $\boldsymbol\delta_\mu=(\boldsymbol\delta_{\mu_1}^\top,\boldsymbol\delta_{\mu_2}^\top)^\top$. The basis function matrix $\boldsymbol\Phi_\sigma$ is defined similarly, with different block-diagonal entries $\boldsymbol\Phi_{\sigma_1}$ and $\boldsymbol\Phi_{\sigma_2}$, and different basis coefficients $\boldsymbol\delta_\sigma=(\boldsymbol\delta_{\sigma_1}^\top,\boldsymbol\delta_{\sigma_2}^\top)^\top$. A tabulated summary of the dimensions and notation can be found in the supplement.

\paragraph{Scalability improvements}
This framework reduces the computational burden from inverting two dense covariance matrices of dimension \( n + m \), which incurs a cubic computational cost with respect to the number of spatial locations, to performing multiplications involving an \( n \times p \) matrix with a \( p \)-vector, an \( m \times p \) matrix with a \( p \)-vector, and an \( m \times q \) matrix with a \( q \)-vector, all of which scale linearly with the number of spatial locations.


\paragraph{Asymmetry and Nonstationarity}

It may be unrealistic to assume that the spatial processes \(\boldsymbol w(\boldsymbol s) = (w_1(\boldsymbol s), \ldots, w_p(\boldsymbol s))\) are independent, as this assumption implies a symmetric cross-covariance, \(\mathsf{Cov}(Z_i(\boldsymbol s), Z_j(\boldsymbol s')) = \mathsf{Cov}(Z_i(\boldsymbol s'), Z_j(\boldsymbol s))\). To explore this, let:
\[
\boldsymbol w(\boldsymbol s) = \boldsymbol A \begin{pmatrix} v_1(\boldsymbol s) \\ v_2(\boldsymbol s) \end{pmatrix} \approx \boldsymbol A \begin{pmatrix} \boldsymbol\Phi_1^\top(\boldsymbol s)\boldsymbol\delta_1 \\ \boldsymbol\Phi_2^\top(\boldsymbol s)\boldsymbol\delta_2 \end{pmatrix},
\]

\noindent with the cross-covariance of \(\boldsymbol v(\boldsymbol s)\) given by:
\[
\boldsymbol C_v(\boldsymbol s, \boldsymbol t) = \begin{pmatrix} \boldsymbol\Phi_1^\top(\boldsymbol s)\boldsymbol\Sigma_{\boldsymbol\delta_1}\boldsymbol\Phi_1(\boldsymbol t) & 0 \\ 0 & \boldsymbol\Phi_2^\top(\boldsymbol s)\boldsymbol\Sigma_{\boldsymbol\delta_2}\boldsymbol\Phi_2(\boldsymbol t) \end{pmatrix}.
\]

\noindent Thus, the cross-covariance of \(\boldsymbol w(\boldsymbol s)\) becomes $\boldsymbol C_w(\boldsymbol s, \boldsymbol t) = \boldsymbol A \boldsymbol C_v(\boldsymbol s, \boldsymbol t) \boldsymbol A^\top$, which results in the symmetry $\boldsymbol C_w(\boldsymbol s, \boldsymbol t)=\boldsymbol C_w^\top(\boldsymbol s, \boldsymbol t)$. Further details can be found in the supplement.

However, environmental, atmospheric, and geophysical processes are often influenced by prevailing winds or ocean currents, which conflict with the assumption of symmetry \citep{apanasovich2010cross}. To account for asymmetry in cross-covariance while maintaining nonstationarity, one can model dependence across basis functions. The cross-covariance of \(\boldsymbol v(\boldsymbol s)\) becomes $\mathsf{Cov}(v_1(\boldsymbol s),v_2(\boldsymbol t))=\boldsymbol\Phi_1^\top(\boldsymbol s)\boldsymbol\Sigma_{\boldsymbol\delta_1,\boldsymbol\delta_2}\boldsymbol\Phi_2(\boldsymbol t)$, resulting in a cross-covariance structure for \(\boldsymbol v(\boldsymbol s)\):

\[
\boldsymbol C_v(\boldsymbol s,\boldsymbol t) = \begin{pmatrix}\boldsymbol\Phi_1^\top(\boldsymbol s)\boldsymbol\Sigma_{\boldsymbol\delta_1}\boldsymbol\Phi_1(\boldsymbol t) & \boldsymbol\Phi_1^\top(\boldsymbol s)\boldsymbol\Sigma_{\boldsymbol\delta_1,\boldsymbol\delta_2}\boldsymbol\Phi_2(\boldsymbol t) \\ \boldsymbol\Phi_2^\top(\boldsymbol s)\boldsymbol\Sigma_{\boldsymbol\delta_2,\boldsymbol\delta_1}\boldsymbol\Phi_1(\boldsymbol t) & \boldsymbol\Phi_2^\top(\boldsymbol s) \boldsymbol\Sigma_{\boldsymbol\delta_2}\boldsymbol\Phi_2(\boldsymbol t)\end{pmatrix}.
\]

\noindent For \(\boldsymbol w(\boldsymbol s)\), the cross-covariance is $\boldsymbol C_w(\boldsymbol s, \boldsymbol t) = \boldsymbol A \boldsymbol C_v(\boldsymbol s, \boldsymbol t)\boldsymbol A^\top$, which results in an asymmetric cross-covariance structure. Additionally, the cross-covariance is not solely a function of the distance between locations, indicating a nonstationary spatial structure.

\paragraph{Dependence Across Basis Functions: Maintaining positive semidefiniteness}
Constructing a joint covariance structure \(\boldsymbol\Sigma_{\boldsymbol\delta}\) for basis coefficients that allows for asymmetric cross-covariances while ensuring nonnegative definiteness is challenging. To achieve this, we can use the framework by \cite{oliver2003gaussian} to generate a positive semidefinite covariance matrix by modeling dependence across basis functions. The process is defined as: 
$\delta_1 = \mu_1 + L_1 Z_1, \quad \delta_2 = \mu_2 + L_2(\rho Z_1 + \sqrt{1 - \rho^2} Z_2)$, where \(L_1\) and \(L_2\) are square roots of covariance matrices \(C_{11}\) and \(C_{22}\), respectively, and \(Z_1\), \(Z_2\) are i.i.d. standard normal vectors. The covariances of \(\delta_1\) and \(\delta_2\) are: $\mathsf{Cov}(\delta_1) = L_1 L_1^\top, \quad \mathsf{Cov}(\delta_2) = L_2 L_2^\top$
and the cross-covariance is: $\mathsf{Cov}(\delta_1, \delta_2) = \rho L_1 L_2^\top$. Thus, the joint covariance matrix is:
\[
L L^\top = \begin{pmatrix}
L_1 L_1^\top & \rho L_1 L_2^\top \\
\rho L_2 L_1^\top & L_2 L_2^\top
\end{pmatrix}.
\]

\noindent This structure is positive semidefinite for all \(\rho \geq -1\). Hence, the joint latent spatial processes for the location and scale parameters maintain positive semidefiniteness, ensuring a valid covariance structure for model fitting. Additional details can be found in the supplement.

\subsection{Bayesian Hierarchical Model for JLS-GEV}
\label{mainmodel}

For the shape parameters \(\boldsymbol\xi\), we specify \(\xi_1, \xi_2 \sim \mathcal{N}^+(0, 100)\). We complete the hierarchical model by specifying prior distributions for the cross-correlation parameters: $\rho_{\sigma}\sim\text{Unif}(-1,1)$ and $\rho_{\mu}\sim\text{Unif}(-1,1)$. We assume that the LMC coefficients \(a_{11,\mu}\), \(a_{21,\mu}\), \(a_{22,\mu}\), \(a_{11,\sigma}\), \(a_{21,\sigma}\), and \(a_{22,\sigma}\) have a weakly informative normal prior \(\mathcal{N}(0, 100)\). 

In summary, the Bayesian hierarchical model takes the following form:


\begin{equation}\label{EQ:Hier_Model}
\begin{aligned}
&\textbf{Data Model: } &  \boldsymbol Z(\boldsymbol s)\sim\text{GEV}(\boldsymbol\mu(\boldsymbol s),\boldsymbol\sigma(\boldsymbol s),\boldsymbol\xi) \\
& &\boldsymbol \mu=\boldsymbol\mu_{0}+\boldsymbol A_{\mu}^*\boldsymbol\Phi_{\mu}\boldsymbol\delta_{\mu}\\
& & \boldsymbol \sigma=\exp(\boldsymbol\sigma_{0}+\boldsymbol A_{\sigma}^*\boldsymbol\Phi_{\sigma}\boldsymbol\delta_{\sigma})\\
&\textbf{Process Model: } &(\boldsymbol\delta_{\mu_1}^\top,\boldsymbol\delta_{\mu_2}^\top)^\top)\sim\mathcal N\left(0,\begin{pmatrix}
L_{\mu_1} L_{\mu_1}^\top & \rho_{\mu} L_{\mu_1} L_{\mu_2}^\top\\
\rho_{\mu} L_{\mu_2} L_{\mu_1}^\top & L_{\mu_2} L_{\mu_2}^\top
\end{pmatrix}\right)\\
& &(\boldsymbol\delta_{\sigma_1}^\top,\boldsymbol\delta_{\sigma_2}^\top)^\top)\sim\mathcal N\left(0,\begin{pmatrix}
L_{\sigma_1} L_{\sigma_1}^\top & \rho_{\sigma} L_{\sigma_1} L_{\sigma_2}^\top\\
\rho_{\sigma} L_{\sigma_2} L_{\sigma_1}^\top & L_{\sigma_2} L_{\sigma_2}^\top
\end{pmatrix}\right)\\
&\textbf{Parameter Model: } & a_{11,\mu}\sim p(a_{11,\mu}),a_{21,\mu}\sim p(a_{21,\mu}),a_{22,\mu}\sim p(a_{22,\mu})\\
&&a_{11,\sigma}\sim p(a_{11,\sigma}),a_{21,\sigma}\sim p(a_{21,\sigma}),a_{22,\sigma}\sim p(a_{22,\sigma})\\
&&\boldsymbol\xi\sim p(\boldsymbol\xi),\rho_{\sigma}\sim p(\rho_{\sigma}),\rho_{\mu}\sim p(\rho_{\mu})\\
\end{aligned}
\end{equation}

\noindent with prior distributions $p(a_{11,\mu})$, $p(a_{21,\mu})$, $p(a_{22,\mu})$, $p(a_{11,\sigma})$, $p(a_{21,\sigma})$, $p(a_{22,\sigma})$, $p(\boldsymbol\xi)$, $p(\rho_{\sigma})$, and $p(\rho_{\mu})$ specified by the practitioner. 

\section{Simulation Study}
\label{sec:ch4_SimulationStudy}

In this section, we demonstrate the proposed approach through an extensive simulation study that considers multiple dependence structures and missing data levels.  Our study focuses on issues motivated by our application to jointly model extreme precipitation and temperature.  To benchmark performance, we compare our Joint Latent Spatial GEV (JLS-GEV) approach to baseline models which assume a symmetric cross-covariance structure.

\subsection{Study Design}

Let \(\boldsymbol{s}_i \in \mathcal{D} = [0, 5]^2 \subset \mathbb{R}^2\) for \(i = 1, \ldots, N\) denote the spatial locations, and let \(\mathcal{S} = (\boldsymbol{s}_1, \ldots, \boldsymbol{s}_N)\). At these locations, let \(\boldsymbol{Z}_1 = (\boldsymbol{Z}_1(\boldsymbol{s}_1)^\top, \ldots, \boldsymbol{Z}_1(\boldsymbol{s}_N)^\top)^\top\) and \(\boldsymbol{Z}_2 = (\boldsymbol{Z}_2(\boldsymbol{s}_1)^\top, \ldots, \boldsymbol{Z}_2(\boldsymbol{s}_N)^\top)^\top\) represent the response variables for each output. For each location, we generate 10 replicates. The data is split into a training set of \(N_O = 5{,}000\) spatial locations for model fitting and a disjoint holdout dataset of \(N_P = 5{,}000\) locations for validation. The spatial random effects for both processes are generated using a Matérn covariance function with spatial range parameter \(\rho = 1\) and partial sill parameter \(\sigma^2 = 1\).

We evaluate \(2 \times 2 \times 5 = 20\) data generation scenarios to compare the JLS-GEV model with baseline models under various conditions. First, we consider scenarios where: (i) only location parameters vary spatially; and (ii) both location and scale parameters vary spatially. Second, we generate data under two cross-covariance conditions: (i) the latent spatial surfaces are negatively correlated with an asymmetric cross-covariance; and (ii) the surfaces are negatively correlated with a symmetric cross-covariance. Third, we consider the issue of the second process having considerably less observations by examining five cases where the proportion of observed $Z_2(\cdot)$ locations is $\{1/50,1/35,1/25,1/15,1/10\}$. The first scenario reflects our motivating application, where the number of temperature locations is about 1/50 of the precipitation locations. A summary table of the simulation study design and data generation details is provided in the supplement.


To assess predictive performance, we calculate the root mean squared error for the 10-year return level surface. Additional metrics, including predictions for the location and scale surfaces, tail area predictions, the continuous ranked probability score (CRPS), and log scores (LogS), are detailed in the supplementary material. We evaluate computational cost via the average computation time for each method in seconds.


\subsection{Results}

We begin by reviewing the simulation results in the most challenging data generation scenario where: a) both the location and scale parameters are spatially varying, while the shape parameters remain fixed but unknown; and b) the two processes are negatively correlated with an asymmetric cross-covariance structure. Additional simulation study results can be found in the supplement. After presenting the JLS-GEV model fit in the most challenging scenario, we provide quantitative comparisons to baseline models.  

\begin{figure}[H]
\begin{center}
\begin{tabular}{cc}
\includegraphics[width=58mm]{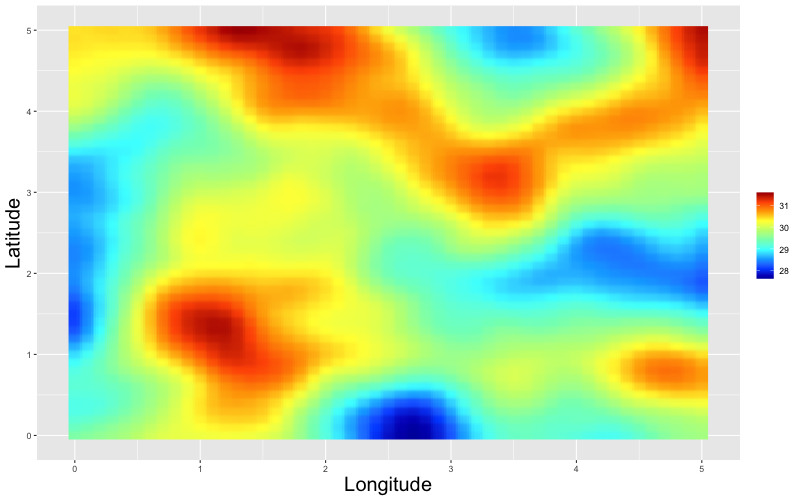} &   \includegraphics[width=58mm]{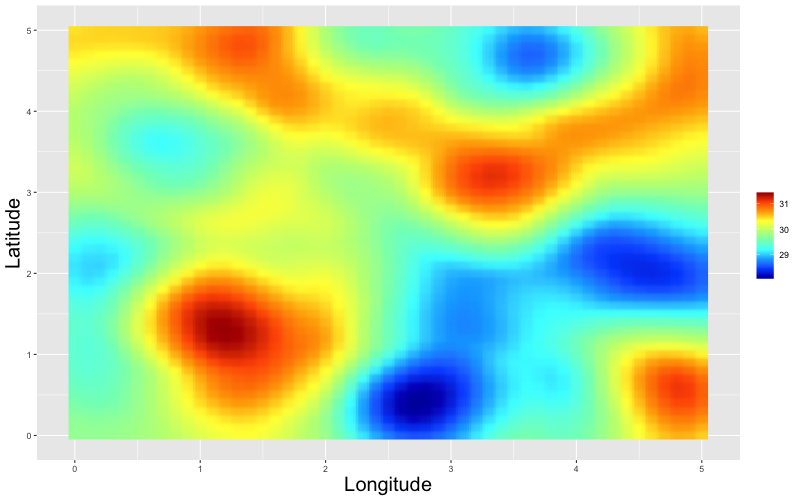} \\
(a) True $Z_1$ Location Surface & (b) Predictive Location Surface for $Z_1$  \\[8pt]
\includegraphics[width=58mm]{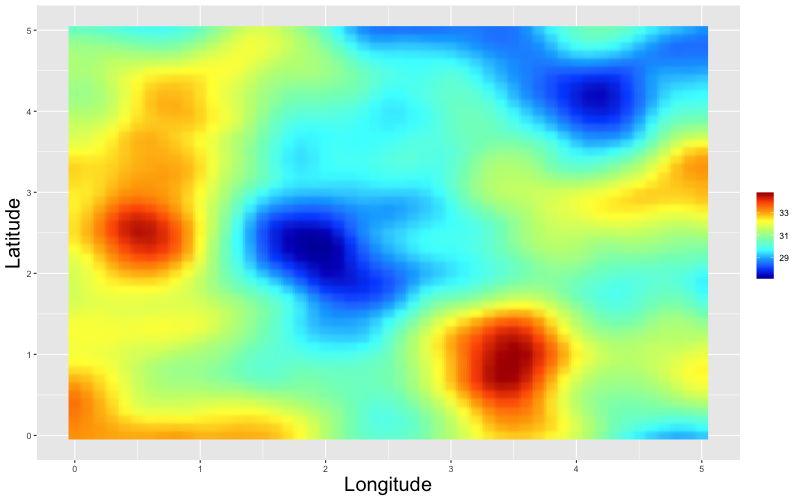} &   \includegraphics[width=58mm]{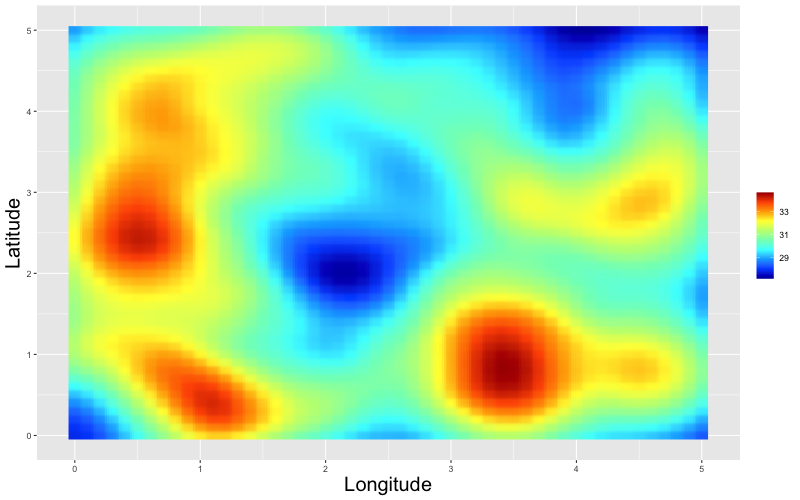} \\
(c) True $Z_2$ Location Surface & (d) Predictive Location Surface for $Z_2$ \\[8pt]
\includegraphics[width=58mm]{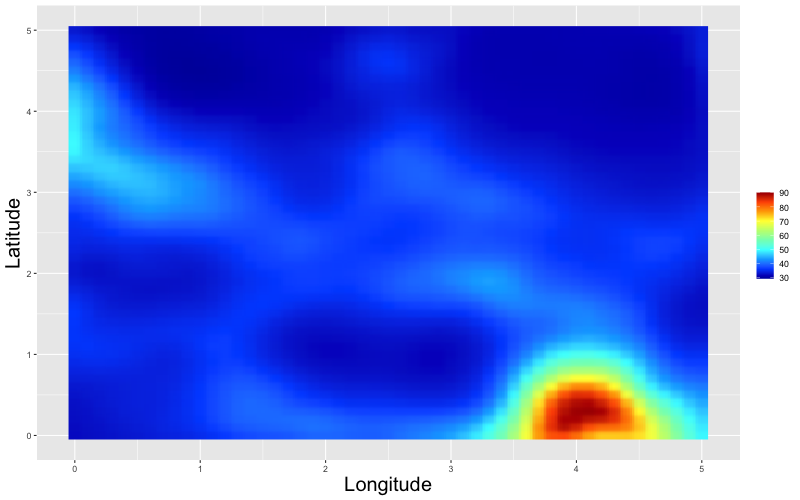} &   \includegraphics[width=58mm]{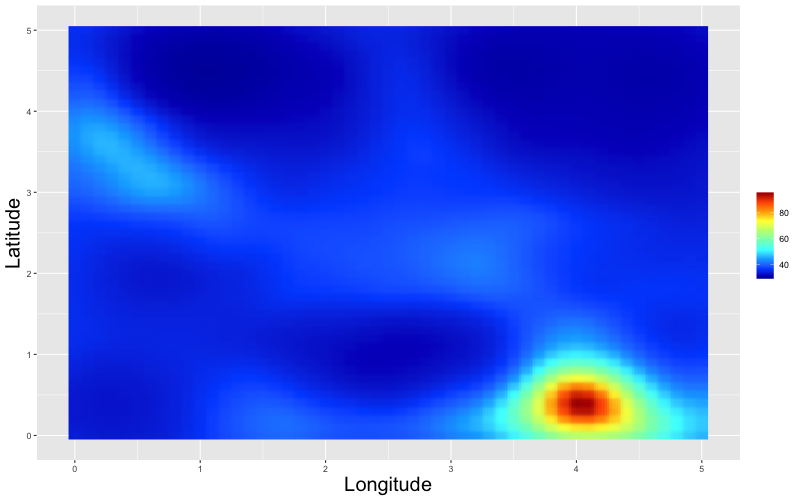} \\
(e) True $Z_1$ Return Level Surface & (f) Predicted $Z_1$ Return Level Surface \\[8pt]
\includegraphics[width=58mm]{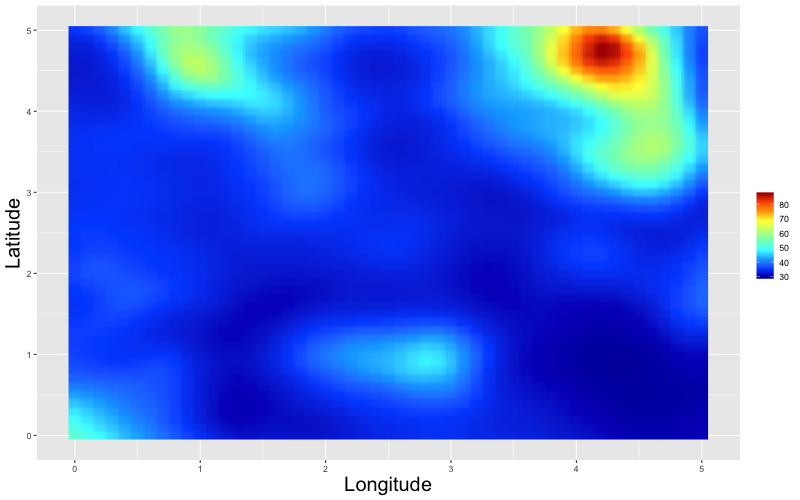} &   \includegraphics[width=58mm]{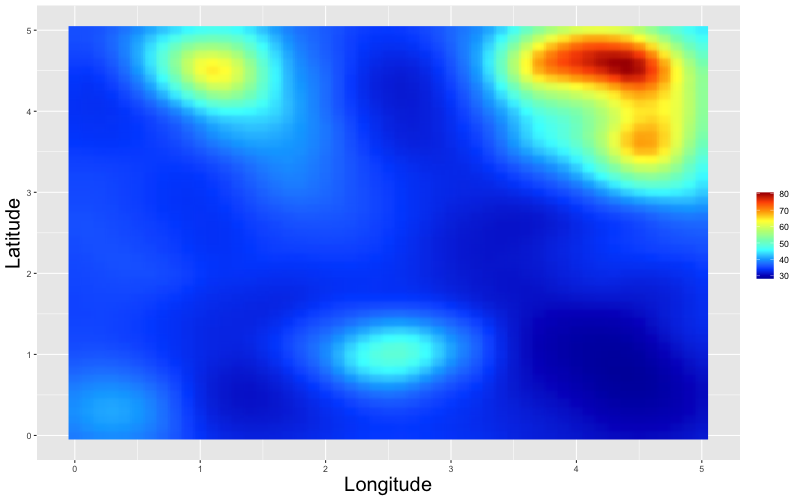} \\
(g) True $Z_2$ Return Level Surface & (h) Predicted $Z_2$ Return Level Surface \\
\end{tabular}
\caption{Comparison of true (left column) and predictive (right column) surfaces for $Z_1$ and $Z_2$. Panels (a) and (b) show the true and predicted location surfaces for $Z_1$, while panels (c) and (d) show the corresponding surfaces for $Z_2$. Panels (e) and (f) display the true and predicted 100-year return level surfaces for $Z_1$, with panels (g) and (h) showing the same for $Z_2$.
}
\label{Fig1:LocScale_mu}
\end{center}
\end{figure}

\paragraph{Example of JLS-GEV model fit}
We present an illustration of the JLS-GEV model fit for the case where only 1/50 of the \(Z_2(\cdot)\) observations are available, with spatially varying location and scale parameters, and an asymmetric cross-covariance structure. Figure \ref{Fig1:LocScale_mu} shows the true and predicted location and return level surfaces for both the \(Z_1(\cdot)\) and \(Z_2(\cdot)\) processes. The JLS-GEV model effectively uses information from the \(Z_1\) process to provide accurate estimates of the spatially varying location and 100-year return level surfaces.

\paragraph{Quantitative evaluation}
Table \ref{Tab1:SimulationStudy_LocScale_Asym_RL10} presents the prediction results for the 10-year return levels. Prediction results for the location and scale surfaces can be found in the supplement. For each metric, the joint asymmetric model consistently yields more accurate predictions than the joint and independent symmetric models for the \(Z_2\) process, with similar results for the $Z_1$ process. The performance gains are more pronounced for the \(Z_2\) process when the proportion of missing \(Z_2\) observations is larger. Supplementary results indicate that the proposed method maintains comparable performance to baseline models even under model misspecification (i.e., when the cross-covariance structure is symmetric). 

Overall, these quantitative results support the qualitative observation from the challenging example presented: the JLS-GEV model effectively leverages information from the \(Z_1\) process to support understanding of the \(Z_2\) process. For our motivating application with roughly a 50-fold reduction in the number of temperature locations as compared to precipitation locations, this simulation study shows potential for precipitation data to strengthen our joint model overall.   

\begin{table}[H]
\singlespacing
\begin{center} 
\begin{tabular}{ccccccccc} \toprule 
&\multicolumn{2}{c}{Joint Asym} && \multicolumn{2}{c}{Joint Sym}&& \multicolumn{2}{c}{Ind Sym}\\
\cmidrule{2-3}\cmidrule{5-6}\cmidrule{8-9}
\makecell{Percentage\\Observed}&$\text{rMSE}_1$ & $\text{rMSE}_2$ && $\text{rMSE}_1$&$\text{rMSE}_2$&& $\text{rMSE}_1$&$\text{rMSE}_2$\\
\midrule
1/10& 1.609&2.592&& 1.545& 3.165&& 1.551& 3.590\\
1/15& 1.654&2.605&& 1.595& 3.568&& 1.583& 3.883\\
1/25& 1.700&3.008&& 1.648& 4.173&& 1.654& 4.661\\
1/35& 1.687&3.247&& 1.600& 3.964&& 1.610& 5.169\\
1/50& 1.649&3.309&& 1.578& 4.425&& 1.589& 5.755\\
\bottomrule \end{tabular}
\end{center}
\caption{Average $\text{rMSE}_{RL_{10,1}}$ and $\text{rMSE}_{RL_{10,2}}$ for the simulation study where both the location and scale parameters are spatially varying. Results are presented for various choices of the percentage of observed $Z_2$ realizations.}
\label{Tab1:SimulationStudy_LocScale_Asym_RL10}
\vspace*{3\baselineskip}
\end{table}

\paragraph{Computational evaluation}

Table \ref{Tab1:SimulationStudy_CompTimes} displays the average computation time for each method in seconds. Additionally, we report the computation time for one implementation of the traditional linear model of coregionalization. While the proposed joint asymmetric model is more costly to fit than the joint and independent symmetric model counterparts, the difference in computation time is relatively small. Each method is significantly faster than a traditional coregionalization model, which would be prohibitive to fit with 5,000 spatial locations.

\begin{table}[H]
\singlespacing
\begin{center} 
\begin{tabular}{ccccc} \toprule 
\makecell{Percentage\\Observed}&\makecell{Joint\\Asym}& \makecell{Joint\\Sym}&\makecell{Independent\\Sym (baseline)}&Full GP (baseline) \\
\midrule
1/10& 2070&1918&1971&1,436,150\\
1/15& 1873&1550&1567&591,600\\
1/25& 1305&1195&1210&234,750\\
1/35& 1436&1243&1268&191,600\\
1/50& 1213&1198&1184&179,450\\
\bottomrule \end{tabular}
\end{center}
\caption{Average computation time (seconds) for the simulation study with spatially varying location parameter. Results are presented for various choices of the percentage of observed $Z_2$ realizations.} 
\label{Tab1:SimulationStudy_CompTimes}
\end{table}

\section{Application: Extreme Precipitation and Temperature in the Southeastern US}
\label{sec:ch4_Applications}

In this section, we apply the JLS-GEV model to analyze the extreme temperature and precipitation data presented in Section \ref{sec:motivation}. To evaluate predictive performance, we compute (1) the median CRPS; (2) the median LogS; (3) the median CRPS in the tail; (4) the median LogS in the tail; and (5) the rMSE between the true and predicted observations in the upper 5\% tail. Here, we consider the joint model where only the location parameter varies spatially; complete results using various other model settings are provided in the supplement. To benchmark performance, we compare against a model that treats each output independently.

Figure \ref{fig:RL_precip} and \ref{fig:RL_temp} below show the predicted 10-year return levels for precipitation and temperature, respectively. Precipitation return levels are highest in Alabama, Mississippi, and Tennessee. In contrast, southern Florida has the lowest predicted return levels. Temperature return levels are largest in Florida. However, eastern Tennessee and western North Carolina have the smallest predicted temperature return levels. These findings align with previous studies that identified similar regional patterns for precipitation \citep{brown2019climatology} and temperature \citep{diem2017heat}. Table \ref{Tab1:realworld_app} presents the results for both the proposed joint and independent model. While both models perform comparably, they exhibit significant scalability improvements over a traditional coregionalization model (without dimension-reduction), which would be infeasible to apply to a dataset of this magnitude. While this implementation yields similar results across all three models, we find that this is an illuminating demonstration of the scalable JLS-GEV methodology for multivariate spatial extremes. 

More pronounced differences between the models emerge when the data is restricted to temperature and precipitation measurements from the summer months (June, July, August). By focusing on this subset, we can isolate the extreme right-tailed values (e.g., heavier rainfall and higher temperatures) characteristic of summer. Predictive results show that the joint asymmetric model outperforms the joint symmetric and independent models across all tail-area metrics (see Table 38 in the supplement). The computational performance of the models remains consistent with their behavior on the full dataset of temperature and precipitation.

\begin{figure}[h]
    \centering
    \begin{subfigure}[t]{0.45\textwidth}
        \centering
        \includegraphics[width=1\linewidth]{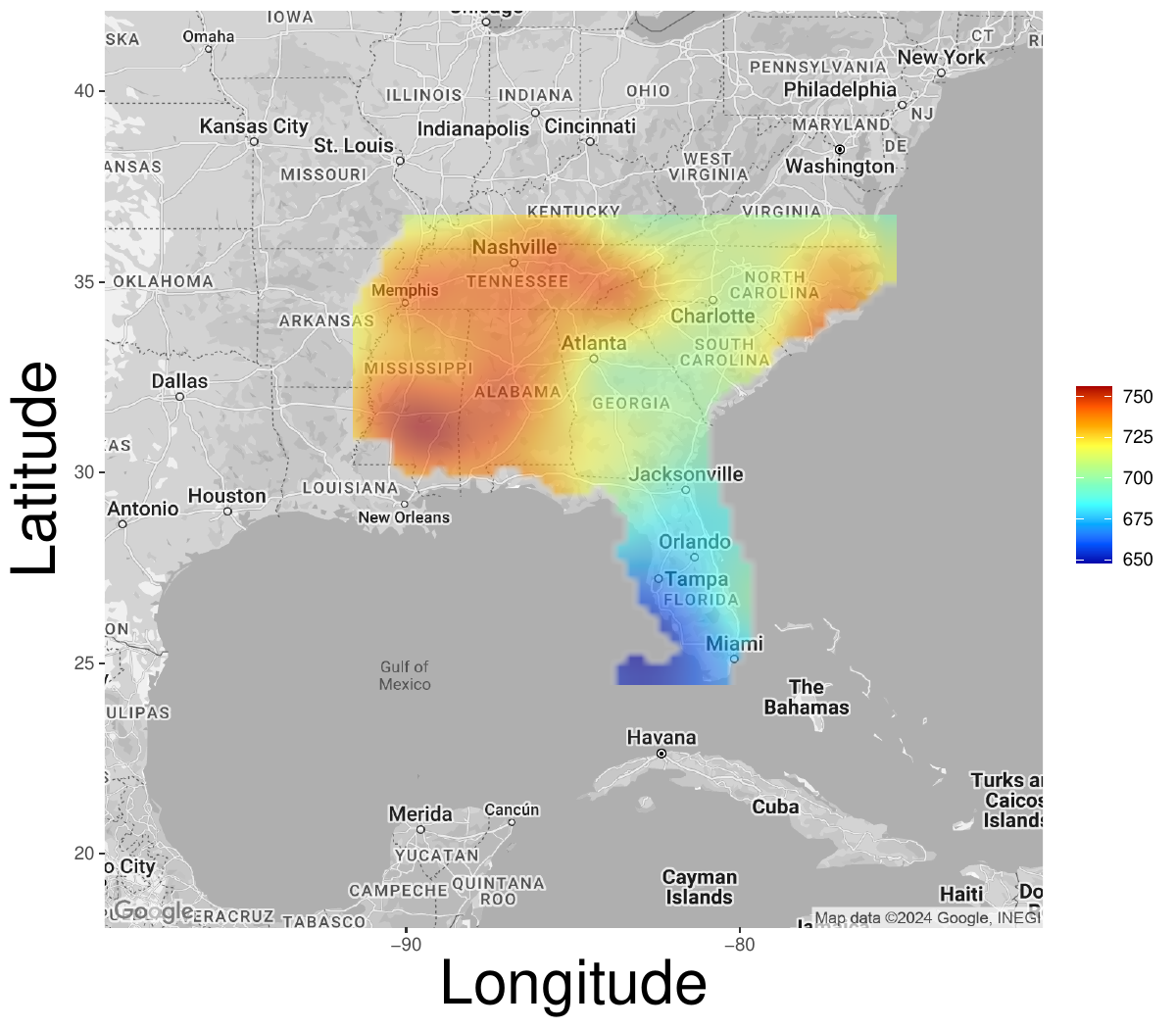} 
        \caption{Precipitation Return Levels} \label{fig:RL_precip}
    \end{subfigure}
    \hfill
    \begin{subfigure}[t]{0.45\textwidth}
        \centering
        \includegraphics[width=1\linewidth]{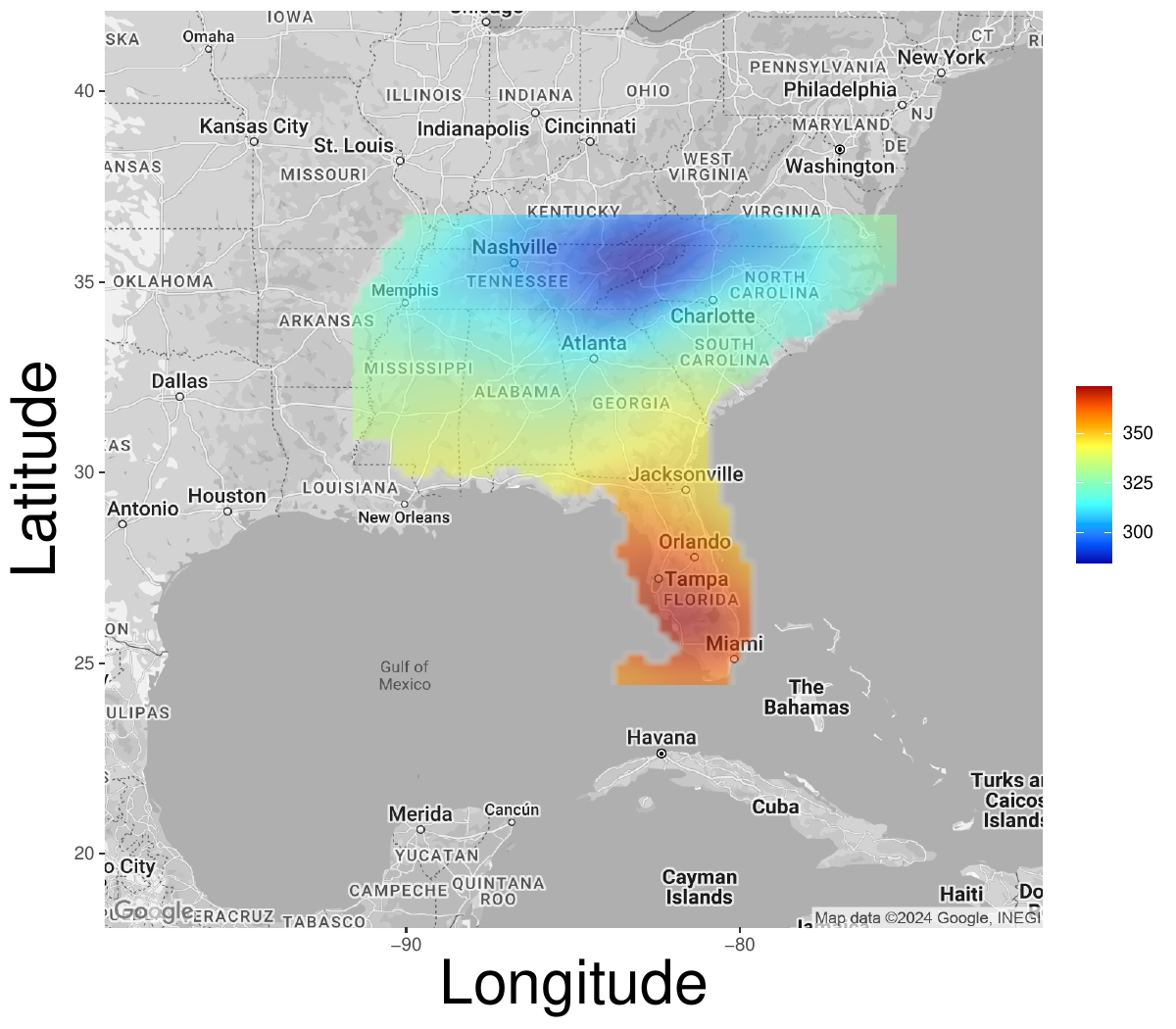} 
        \caption{Temperature Return Levels} \label{fig:RL_temp}
    \end{subfigure}
    \caption{Precipitation 10-year return levels (tenths of millimeters) (left) and temperature 10-year return levels (tenths of degrees Celsius) (right).}
\end{figure}

\begin{table}[h] 
\begin{center} 
\begin{tabular}{ccccccccc} \toprule 
&\multicolumn{2}{c}{Joint Model} && \multicolumn{2}{c}{Joint Sym} &&\multicolumn{2}{c}{Independent}\\
\cmidrule{2-3}\cmidrule{5-6}\cmidrule{8-9}
Metric&Precip. & Temp. && Precip. & Temp. &&Precip. & Temp.\\
\midrule
CRPS &86.27& 22.13&&86.28& 22.15&&86.27&22.07\\
LogS &6.46&5.16&&6.46& 5.17&&6.46&5.16\\
$\text{CRPS}_{\text{tail}}$ &555.30&43.90&&555.54& 44.04&&555.44&43.81\\
$\text{LogS}_{\text{tail}}$ &9.29&5.91&&9.29& 5.91&&9.29&5.91\\
$\text{rMSE}_{\text{tail}}$ &896.85&155.90&&896.95& 155.92&&896.93&155.88\\
\bottomrule \end{tabular}
\end{center}
\caption{Median $\text{CRPS}$, $\text{LogS}$, $\text{CRPS}_{\text{tail}}$, $\text{LogS}_{\text{tail}}$, and $\text{rMSE}_{\text{tail}}$ for the real-world example with spatially varying location parameter. Results are presented for both the joint model and independent model, and for both precipitation and temperature outputs.}
\label{Tab1:realworld_app}
\vspace*{3\baselineskip}
\end{table}

\section{Discussion}
\label{sec:ch4_discussion}

We introduce a new approach (JLS-GEV) for jointly modeling spatial extreme data that is designed to handle large datasets while providing flexibility in the cross-covariance structure. Asymmetry in the cross-covariances is induced by adjusting the dependence among the spatial basis coefficients, which yields an asymmetric cross-covariance structure that maintains positive semi-definiteness. Through a simulation study, we demonstrate that our approach outperforms baseline models that assume symmetric cross-covariances. We apply our method to a real-world dataset, focusing on extreme precipitation and extreme temperature in the southeastern United States. Our JLS-GEV framework is scalable to large, multivariate spatial datasets and can be extended to multivariate models outside of the spatial GEV framework. In particular, the proposed method is well-suited to model high-dimensional environmental datasets while balancing model flexibility with computation efficiency.  We hope that future work will build on these initial results to jointly understand the spatial variation of extremes in interdependent natural processes.

Extensions to the multivariate spatio-temporal setting can enhance our understanding of extreme events by explicitly modeling \(\boldsymbol{v}(\boldsymbol{s},\boldsymbol{t}) = \boldsymbol{A}\boldsymbol{w}(\boldsymbol{s},\boldsymbol{t})\), where the components of \(\boldsymbol{w}\), \(w_l(\boldsymbol{s},\boldsymbol{t})\), represent independent spatio-temporal processes. Similarly, extending the approach to three or more outputs would allow practitioners to study the compounding and cascading behavior of climate and weather hazards. Estimating additional cross-correlation parameters may increase computational costs; however, integrating multiple outputs can significantly enhance analyses of geographically sparse hazards (e.g., extreme temperatures) by ``borrowing strength'' from more densely observed data (e.g., precipitation). This can be achieved using computationally efficient extensions that exploit shared spatial random fields \citep{schliep2013multilevel} with few factor loading weights; thereby bypassing inference of cross-correlation parameters. 

Future research can include a wider range of basis functions (or combinations) including  tensor-product Bernstein polynomial functions over triangular prismatic partitions \citep{yu2022spatiotemporal}, multiquadric radial basis functions, Fourier basis functions \citep{royle2005efficient}, and Gaussian radial basis functions \cite{macdonald2024flexible}, among others. This study examines data from seven states in the southeastern United States, but future studies can expand our methodology to the entire continental United States. To offset some of the computational costs, practitioners may benefit from using distributed computing and locally-partitioned models \citep{lee2023scalable,heaton2017nonstationary}. 



\section*{Supplementary Material}
The supplementary material includes: (1) details on the construction of the bisquare basis functions as well as a visualization of the multi-resolution ``quad-tree" structure; (2) additional evaluation metrics for the simulation study; (3) additional simulation results; (4) proofs of positive semi-definiteness in the joint covariance structure for the basis coefficients; (5) computation times for the simulation study; and (6) details and justification for imposing joint structure on the basis coefficients to account for asymmetry.

\if0\blind
{

\section*{Acknowledgements}\label{Sec:Acknowledgements}
This project was supported by the National Oceanic and Atmospheric Administration (NOAA) (Award Number NA23OAR0210558). This project was also supported by computing resources provided by the Office of Research Computing at George Mason University (\url{https://orc.gmu.edu}) and funded in part by grants from the National Science Foundation (Awards Number 1625039 and 2018631). 

} \fi


\if1\blind
{
\section{Declarations}
\paragraph{Funding and/or Conflicts of interests/Competing interests}
This project was supported by the National Oceanic and Atmospheric Administration (NOAA) (Award Number NA23OAR0210558). All authors declare that they have no conflicts of interest.
} \fi

\bibliographystyle{apalike}
\bibliography{References}
\end{document}


\doublespacing
\maketitle

\section{Proof of Concept}
\label{sec:ch4_proof_concept}

In this section, we provide a proof of concept of our approach. Additional proof of concepts for Gaussian, binary, and count data is provided in the Appendix. Suppose that we have a two dimensional input $\boldsymbol x\in\mathbb R^{N\times 2}$ and a two-dimensional output $\boldsymbol Z\in\{\boldsymbol\mu-\sigma/\xi,\infty\}^{N\times 2}$, where $\boldsymbol\mu$ is the spatially varying location parameter, $\sigma$ is the scale parameter, and $\xi$ is the shape parameter. Figure \ref{fig:y1_true} and Figure \ref{fig:y2_true} below illustrates the location surfaces at the data locations, where ``$Z_1$'' corresponds to the first output dimension and ``$Z_2$'' corresponds to the second output.
\begin{figure}[h]
    \centering
    \begin{subfigure}[t]{0.45\textwidth}
        \centering
        \includegraphics[width=1\linewidth]{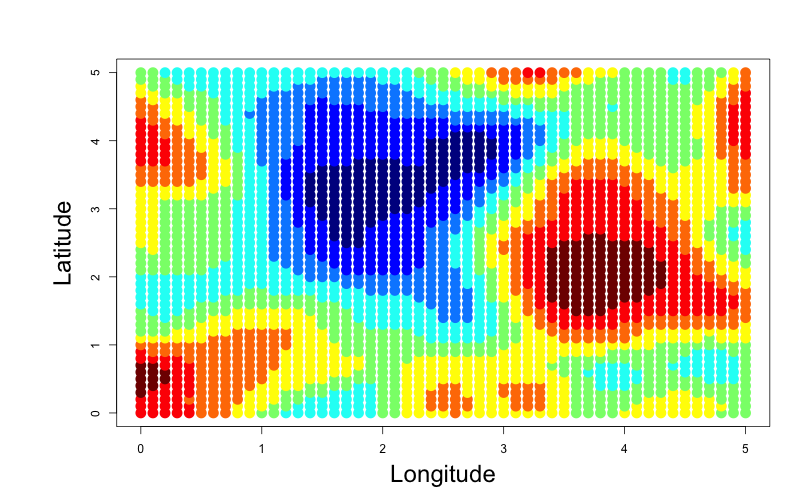} 
        \caption{True $Z_1$ Location Surface} \label{fig:y1_true}
    \end{subfigure}
    \hfill
    \begin{subfigure}[t]{0.45\textwidth}
        \centering
        \includegraphics[width=1\linewidth]{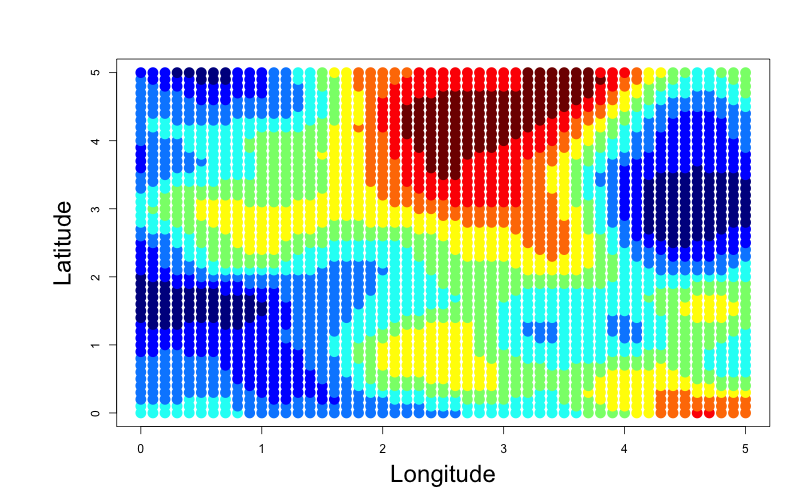} 
        \caption{True $Z_2$ Location Surface} \label{fig:y2_true}
    \end{subfigure}
    \caption{True Location Surfaces\vspace*{3\baselineskip}}
\end{figure}

We can see that the two location surfaces are highly (inversely) correlated with one another. Further, the cross-covariance is asymmetric due to the spatial delay (shift) in the covariance between each of the two processes. This correlation should allow us to share strength across the outputs in our predictions, rather than treating each of the outputs as independent. To show this, let us suppose that we only observe a tenth of the samples for the second output, and we would like to predict the output at the remaining locations. Figure \ref{fig:y1_true_2} and Figure \ref{fig:y2_true_obs} below illustrates the location surfaces at the observed data locations.

\newpage
\begin{figure}[h]
\vspace*{3\baselineskip}
    \centering
    \begin{subfigure}[t]{0.45\textwidth}
        \centering
        \captionsetup{justification=centering}
        \includegraphics[width=1\linewidth]{Figures/Y1_True_GEV_Observed2.png} 
        \caption{True $Z_1$ Location Surface at\\Observed Locations}\label{fig:y1_true_2}
    \end{subfigure}
    \hfill
    \begin{subfigure}[t]{0.45\textwidth}
        \centering
        \captionsetup{justification=centering}
        \includegraphics[width=1\linewidth]{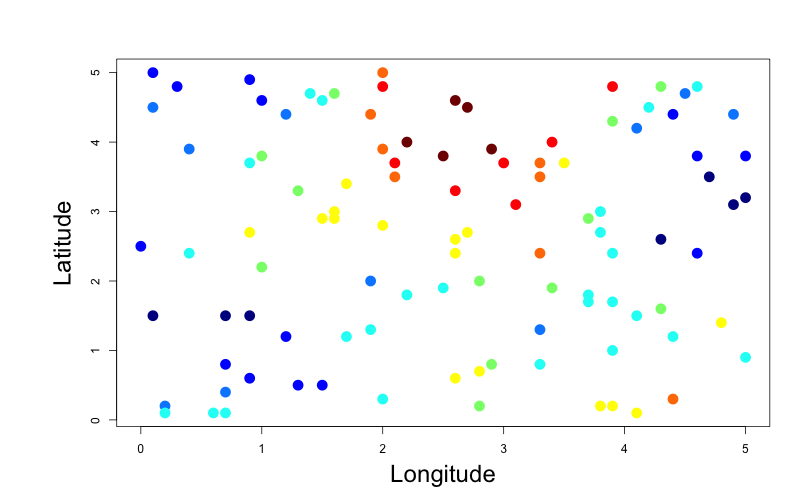}
        \caption{True $Z_2$ Location Surface at\\Observed Locations} \label{fig:y2_true_obs}
    \end{subfigure}
    \caption{True Location Surfaces at Observed Locations\vspace*{3\baselineskip}}
\end{figure}

\begin{figure}[H]
\begin{center}
\begin{tabular}{cc}
\includegraphics[width=70mm]{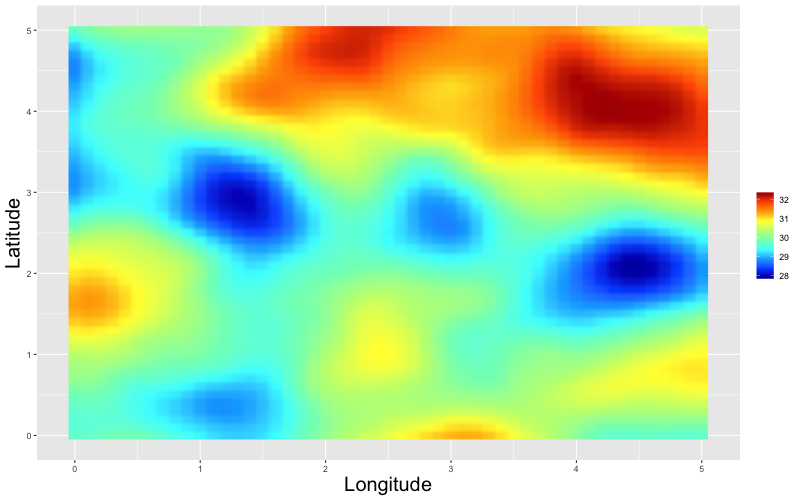} &   \includegraphics[width=70mm]{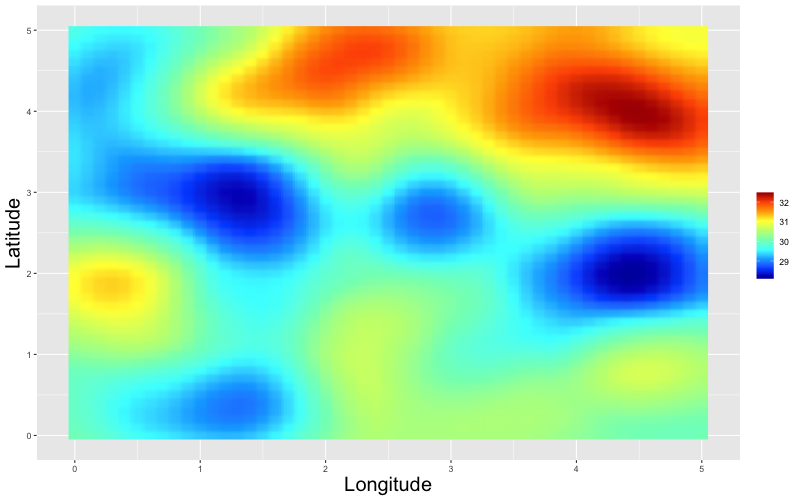} \\
(a) True $Z_1$ Location Surface & \makecell{(b) Predictive Location Surface\\ for $Z_1$ (Joint Model)} \\[20pt]
\includegraphics[width=70mm]{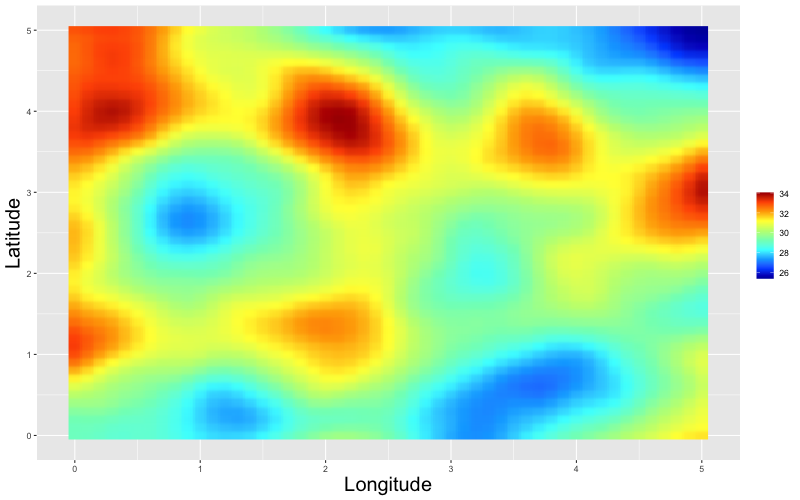} &   \includegraphics[width=70mm]{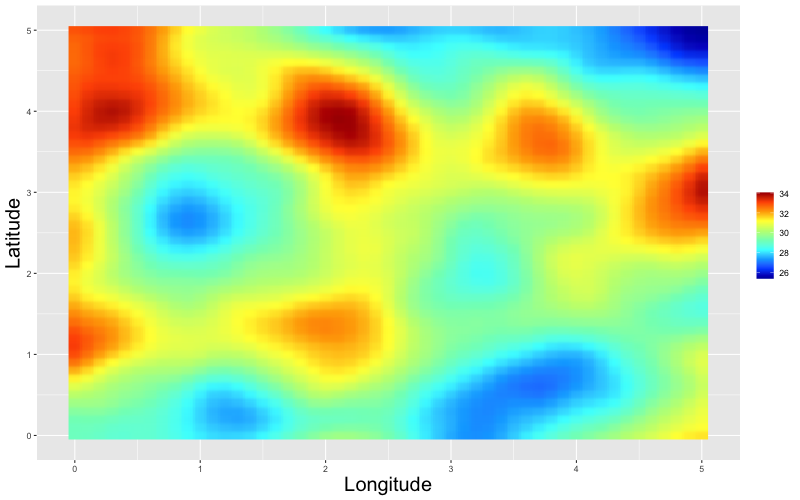} \\
(c) True $Z_2$ Location Surface & \makecell{(d) Predictive Location Surface\\ for $Z_2$ (Joint Model)} \\[20pt]
\includegraphics[width=70mm]{Figures/Y2_True_GEV_Asym2.png} &   \includegraphics[width=70mm]{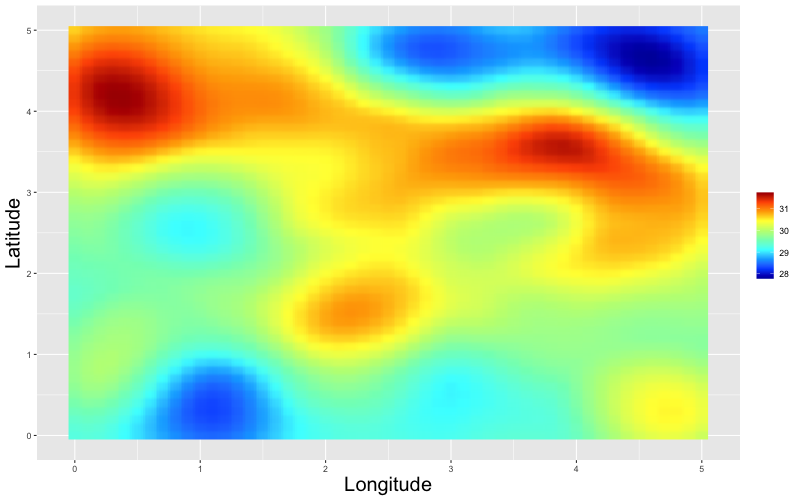} \\
(e) True $Z_2$ Location Surface & \makecell{(f) Predictive Location Surface\\ for $Z_2$ (Independent GPs)} \\[2pt]
\end{tabular}
\caption{True (left column) and predictive (right column) location surfaces
}
\label{Fig1:Joint_Norm_Spatial}
\end{center}
\end{figure}

Figure \ref{Fig1:Joint_Norm_Spatial}d below displays the predictive location surface for $Z_2$ using the joint model. Figure \ref{Fig1:Joint_Norm_Spatial}f below displays the predictive location surface for $Z_2$ using two independent latent GP models for $Z_1$ and $Z_2$. We can see that the joint model is able to leverage information from $Z_1$ to provide accurate predictions for $Z_2$. On the other hand, the single-output GP seems to oversmooth the spatial process.

\section{Bisquare Basis Functions}
\setcounter{equation}{0}

\begin{figure}[ht]
\centering
\includegraphics[scale=.4]{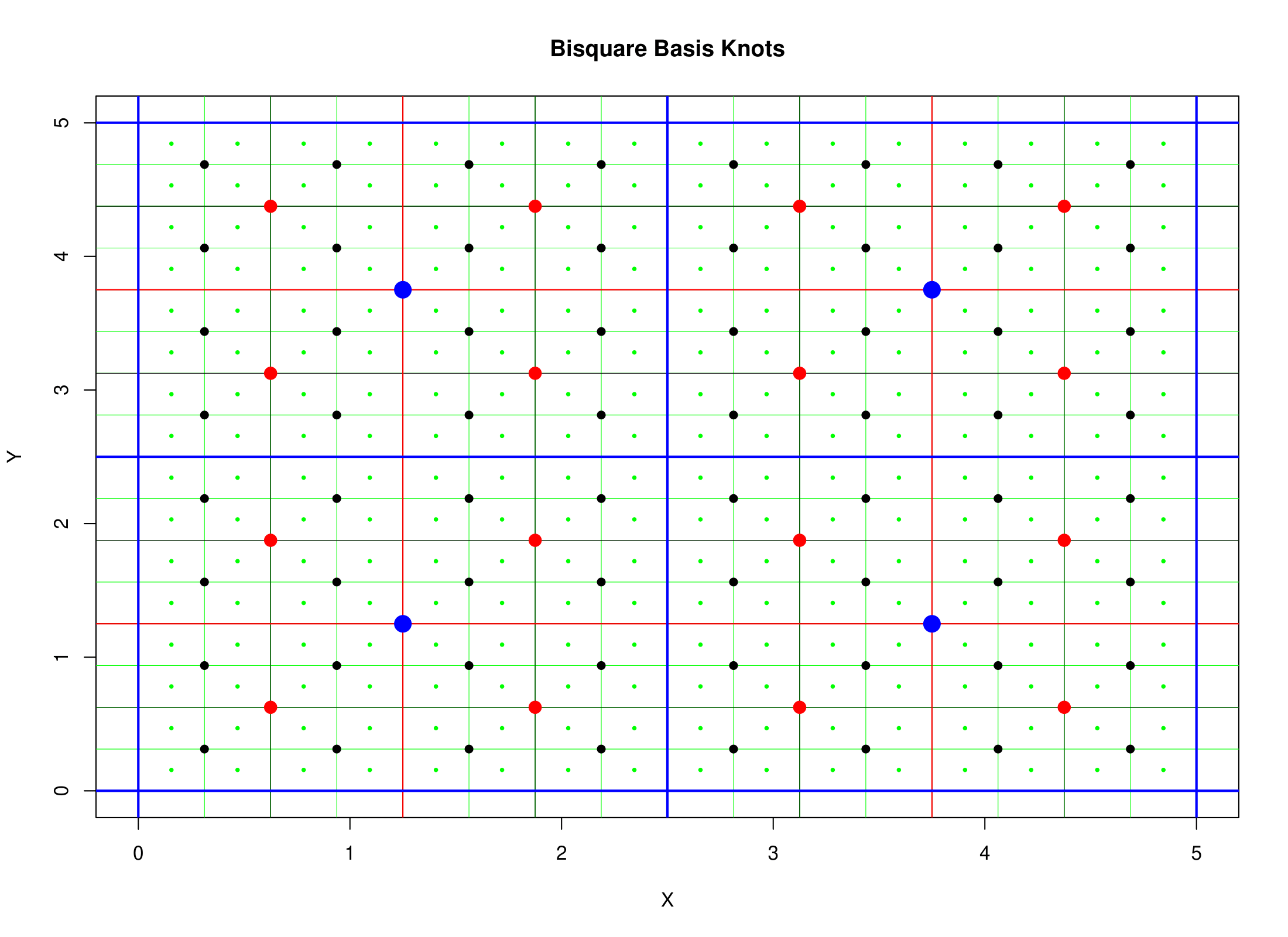}
\caption{Illustration of multi-resolution quad-tree structure}
\label{Fig1:bisquare}
\end{figure}

We employ the bisquare basis functions from \citep{sengupta2013hierarchical,cressie2008fixed} which take the form:
$$\Phi_m(\boldsymbol s)=\left\{1-\left(\frac{||\boldsymbol s-\boldsymbol u_m||}{\gamma}\right)^2\right\}^2\mathds{1}\left(||\boldsymbol s-\boldsymbol u_m||<\gamma\right),$$
\noindent where $\boldsymbol u_m$ is the center of basis function $m$ and $\mathds 1(\cdot)$ is an indicator function. The knots associated with each basis function are constructed according to a multi-resolution ``quad-tree'' structure such that the knots associated with different resolutions do not overlap. In particular, we use three resolutions, where there are four knot locations $\boldsymbol u_1,\ldots,\boldsymbol u_4$ for the first resolution, 16 knots $\boldsymbol u_5,\ldots,\boldsymbol u_{20}$ for the second resolution, and 64 knots $\boldsymbol u_{21},\ldots,\boldsymbol u_{84}$ for the third resolution. An illustration of the three resolutions of knot locations is provided in Figure \ref{Fig1:bisquare}. The bandwidth $\gamma$ for a specific resolution from \cite{cressie2008fixed} is given by $\gamma = 1.5\times\text{ minimum distance between knot locations}$.
We employ the multi-resolution bisquare basis functions into the Bayesian hierarchical framework for spatial generalized linear mixed models (SLGMMs) as follows:

$$\begin{aligned}
&\textbf{Data Model: } &z(\boldsymbol s_i)\mid\eta(\boldsymbol s_i)\sim F(\eta(\boldsymbol s_i)) \\
& &g(\mathbb E[\boldsymbol z\mid\boldsymbol\beta,\boldsymbol\delta]):=\boldsymbol\eta=\boldsymbol X\boldsymbol\beta+\boldsymbol\Phi\boldsymbol\delta\\
&\textbf{Process Model: } &\boldsymbol\delta \mid\tau^2\sim\mathcal N(\boldsymbol 0, \tau^2\mathcal I)\\
&\textbf{Parameter Model: } & \boldsymbol\beta\sim p(\boldsymbol\beta), \tau^2\sim p(\tau^2),
\end{aligned}$$

\noindent where $\mathcal I$ denotes the identity matrix and $\boldsymbol\Phi$ denotes the matrix of the multi-resolution bisquare basis functions. The hierarchical model is completed by assigning prior distributions for the model parameters $\boldsymbol\beta$ and $\tau^2$. In the simulation study, we use the following prior distributions for the model parameters: $\boldsymbol\beta\sim\mathcal N(\bm 0,100\mathcal I)$ and $\tau^2\sim\mathsf{IG}(0.5,2000)$.

\section{Evaluation Metrics}

\sloppy For each of the two output variables, we compute (1) the root mean squared error $\text{rMSE}_{\mu}=\left(\frac{1}{N_P}\sum_{i=1}^{N_P}\left(\mu_{i}-\hat{\mu}_{i}\right)^2\right)^{1/2}$ between the true $\mu$ location surface and the predicted $\mu$ location surface; (2) the root mean squared error $\text{rMSE}_{\sigma}=\left(\frac{1}{N_P}\sum_{i=1}^{N_P}\left(\sigma_{i}-\hat{\sigma}_{i}\right)^2\right)^{1/2}$ between the true $\sigma$ scale surface and the predicted $\sigma$ scale surface; (3) the root mean squared error $\text{rMSE}_{RL_{100}}=\left(\frac{1}{N_P}\sum_{i=1}^{N_P}\left(RL_{100,i}-\widehat{RL}_{100,i}\right)^2\right)^{1/2}$ between the true and predicted 100 year return levels; (4) the root mean squared error $\text{rMSE}_{RL_{10}}=\left(\frac{1}{N_P}\sum_{i=1}^{N_P}\left(RL_{10,i}-\widehat{RL}_{10,i}\right)^2\right)^{1/2}$ between the true and predicted 10 year return levels; (5) the root mean squared error $\text{rMSE}_{\text{tail}}=\left(\frac{1}{N_P^{(\text{tail})}}\sum_{i=1}^{N_P^{(\text{tail})}}\left(Z_{i}-\hat{Z}_{i}\right)^2\right)^{1/2}$ between the true and predicted observations in the upper 5\% tail; (6) the median CRPS; (7) the median LogS; (8) the median CRPS in the tail; and (9) the median LogS in the tail. For each metric, we compute the average across 100 simulated datasets. 

The CRPS \citep{matheson1976scoring} is defined in terms of the predictive CDF $F$ and is given by $\text{CRPS}(F,y)=\int_{\mathbb R}(F(z)-\boldsymbol{1}\{y\leq z\})^2 dz$, where $\boldsymbol{1}\{y\leq z\}$ denotes the indicator function which is one if $y\leq z$ and zero otherwise.
In various applications, the forecast distribution of interest, denoted as $F$, often lacks an analytic form and is instead represented solely through a simulated sample $X_1,\ldots, X_m\sim F$. This scenario arises in Bayesian forecasting, where the sample is generated via MCMC algorithms, or in ensemble weather forecasting, where sample values stem from numerical weather prediction models with diverse model physics and/or initial conditions. To evaluate a proper scoring rule, the simulated sample must be transformed into an estimated distribution, denoted as $\hat{F}_m(z)$, which can be assessed at any point $z\in\mathbb R$. For the CRPS, the empirical Cumulative Distribution Function (CDF), $\hat{F}_m(z)=\frac{1}{m}\sum_{i=1}^m\boldsymbol 1\{X_i\leq z\}$, serves as a natural approximation of the predictive CDF. In this context, the CRPS simplifies to $\text{CRPS}(F_m, y) =\frac{1}{m}\sum_{i=1}^m |X_i-y|-\frac{1}{2m^2}\sum_{i=1}^m\sum_{j=1}^m |X_i-X_j|$, enabling direct computation of the CRPS from the simulated sample \citep{grimit2006continuous}. CRPS assesses the discrepancy between the cumulative distribution function of the forecasted probability distribution and the observed outcome. LogS is expressed as $\text{LogS}(F, y) = -\log(f(y))$, where $f(\cdot)$ represents a kernel density estimate of the probability distribution function. Unlike CRPS, computing LogS demands a predictive density, which can be attained through classical nonparametric kernel density estimation to derive an estimator. However, this estimator is valid only under stringent theoretical assumptions and can be fragile in practice. Particularly, if the outcome resides within the tails of the simulated forecast distribution, the estimated score may exhibit high sensitivity to the choice of the bandwidth tuning parameter. In an MCMC context, opting for a mixture-of-parameters estimator that leverages a simulated sample of parameter draws, rather than draws from the posterior predictive distribution, often proves to be a superior and significantly more efficient choice \citep{kruger2021predictive, jordan2017evaluating}. Therefore, this is the approach we adopt.

\section{Multivariate Spatial Models}
Statistical analyses of multivariate measurements collected at various locations has come to the forefront in spatial statistics \citep{gelfand2021multivariate, schliep2013multilevel}. These data often exhibit dependencies both within measurements at a single location and across different locations. Examples include environmental monitoring of pollutants like ozone and carbon monoxide \citep{barrenetxea2008sensorscope}, studies of plant species assemblages influenced by environmental factors \citep{warton2015model}, and atmospheric modeling with measurements such as temperature, precipitation, and wind speed \citep{diao2019methods}.
\paragraph{Separable Models} The separable specification \citep{brown1994multivariate,sun1998assessment} is a straightforward method for constructing valid cross-covariance matrices in hierarchical nonstationary multivariate spatial models. Let $\boldsymbol T$ be a $p\times p$ positive definite matrix representing the covariance matrix of $\boldsymbol Y(\boldsymbol s)$. Using a valid univariate correlation function $\rho$ that decreases with distance, the cross-covariance function can be defined as $\boldsymbol C(\boldsymbol s,\boldsymbol s')=\rho(\boldsymbol s,\boldsymbol s')\boldsymbol T$. $\boldsymbol T$ captures local associations while $\rho$ governs spatial associations. The covariance matrix for $\boldsymbol Y=(\boldsymbol Y(\boldsymbol s_1),\ldots,\boldsymbol Y(\boldsymbol s_n))^\top$ becomes $\boldsymbol\Sigma_{\boldsymbol Y}=\boldsymbol R\otimes\boldsymbol T$, where $\boldsymbol R$ is an $n\times n$ matrix with entries $(R)_{i,j}=\rho(\boldsymbol s_i,\boldsymbol s_j)$, and $\otimes$ denotes the Kronecker product. This formulation ensures that $\boldsymbol\Sigma_{\boldsymbol Y}$ is positive definite, as both $\boldsymbol R$ and $\boldsymbol T$ are positive definite. It also simplifies computations, as $|\boldsymbol\Sigma_{\boldsymbol Y}|=|\boldsymbol R|^p |\boldsymbol T|^n$ and $\boldsymbol\Sigma_{\boldsymbol Y}^{-1}=\boldsymbol R^{-1}\otimes\boldsymbol T^{-1}$, allowing for efficient updates using smaller matrices. Separable models are limited such that the cross-covariances remain constant across locations and each component of $\boldsymbol w(\boldsymbol s)$ shares the same spatial correlation structure \cite{gelfand2021multivariate}.



\paragraph{Kernel Convolution Methods} 
Moving average methods, also known as kernel convolution, are commonly used to construct valid stationary cross-covariograms \citep{higdon1999non,paciorek2003nonstationary}. Consider a set of square integrable kernel functions \(k_l(\cdot)\), \(l = 1, \ldots, p\), and a zero-mean unit variance GP \(w(\boldsymbol s)\) with correlation function \(\rho\). The spatial process \(\boldsymbol Y(\boldsymbol s)\) is defined as $Y_l(\boldsymbol s) = \sigma_l \int k_l(\boldsymbol s - \boldsymbol t) w(\boldsymbol t) d\boldsymbol t$,
\noindent resulting in a valid stationary cross-covariance function. A discrete approximation replaces the integral with a sum, but this version loses stationarity. An alternative approach involves spatially varying kernels, where \(k_l(\boldsymbol s - \boldsymbol t)\) is replaced by \(k_l(\boldsymbol s - \boldsymbol t; \boldsymbol s)\). The corresponding cross-covariance matrix is then:
\[
(C(\boldsymbol s, \boldsymbol s'))_{l, l'} = \sigma_l \sigma_{l'} \int k_l(\boldsymbol s - \boldsymbol t; \boldsymbol s) k_{l'}(\boldsymbol s' - \boldsymbol t; \boldsymbol s') d\boldsymbol t.
\]
\noindent The specification of kernels are often left to the practitioner. \cite{higdon1999non} employ Gaussian kernels, while alternatives like Mat\'ern kernels \citep{paciorek2003nonstationary} can alleviate issues with oversmoothing.

\paragraph{Convolution of Covariance Functions Approaches} Cross-covariance functions can be constructed by convolving \(p\)-different stationary one-dimensional covariance functions \citep{majumdar2007multivariate,gaspari1999construction}. The cross-covariance function is defined as \(C_{ij}(\boldsymbol{s}) = \int C_i(\boldsymbol{s} - \boldsymbol{t}) C_j(\boldsymbol{t}) d\boldsymbol{t}\) where \(C_1, \ldots, C_p\) are valid stationary covariance functions. If the $C_i$'s are isotropic, the convolved functions \(C_{ij}\) will also be isotropic \citep{gaspari1999construction}. To obtain a valid cross-correlation function \(\boldsymbol{R}(\boldsymbol{s})\), the correlation functions \(\rho_{ij}(\boldsymbol{s})\) are standardized to achieve unit variance as follows:
\[
\rho_{ij}(\boldsymbol{s}) = \frac{C_{ij}(\boldsymbol{s})}{(C_{ii}(\boldsymbol{0})C_{jj}(\boldsymbol{0}))^{1/2}}.
\]

\noindent The resulting cross-covariance function can be parametrized as \(\boldsymbol{C}_{\sigma} = \boldsymbol{D}_{\sigma}^{1/2} \boldsymbol{R}(\boldsymbol{s}) \boldsymbol{D}_{\sigma}^{1/2}\), where \(\boldsymbol{D}_{\sigma}^{1/2}\) is a diagonal matrix of standard deviations. This method ensures that \(|\rho_{ij}(\boldsymbol{s})| \leq 1\) and provides a consistent interpretation of the correlation structure. While closed-form solutions exist for some cases, Monte Carlo integration is often required to compute the cross-covariances \cite{gelfand2021multivariate}.

\section{Additional Simulation Study Results}

\subsection{Asymmetric Data Generation}

We begin by investigating the simulation results in a scenario where both the location and scale parameters are spatially varying, while the shape parameters remain fixed but unknown. Then we study the case where only the location parameter is spatially varying. We focus on the case where the latent spatial surfaces for each process are negatively correlated with an asymmetric cross-covariance structure.

\noindent\textbf{Spatially Varying Location and Scale Parameters}

Table \ref{Tab1:SimulationStudy_LocScale_Asym_Mu} displays the prediction results for the location surfaces for each process. Table \ref{Tab1:SimulationStudy_LocScale_Asym_Sigma} displays the prediction results for the scale surfaces for each process. Tables \ref{Tab1:SimulationStudy_LocScale_Asym_CRPS} and \ref{Tab1:SimulationStudy_LocScale_Asym_LogS} show the average CRPS and LogS for each process. For each metric, the joint asymmetric model consistently yields more accurate predictions than the joint and independent symmetric models for the $Z_2$ process, with similar results for the $Z_1$ process.

\begin{table}[H]
\singlespacing
\begin{center} 
\begin{tabular}{ccccccccc} \toprule 
&\multicolumn{2}{c}{Joint Asym} && \multicolumn{2}{c}{Joint Sym}&& \multicolumn{2}{c}{Ind Sym}\\
\cmidrule{2-3}\cmidrule{5-6}\cmidrule{8-9}
\makecell{Percentage\\Observed}&$\text{rMSE}_{\mu_1}$ & $\text{rMSE}_{\mu_2}$ && $\text{rMSE}_{\mu_1}$&$\text{rMSE}_{\mu_2}$&& $\text{rMSE}_{\mu_1}$&$\text{rMSE}_{\mu_2}$\\
\midrule
1/10& 0.253&0.653&& 0.252& 0.831&& 0.254& 0.820\\
1/15& 0.262&0.660&& 0.260& 0.924&& 0.255& 0.884\\
1/25& 0.256&0.793&& 0.264& 1.040&& 0.258& 1.037\\
1/35& 0.265&0.909&& 0.254& 1.080&& 0.263& 1.177\\
1/50& 0.262&1.139&& 0.261& 1.189&& 0.254& 1.197\\
\bottomrule \end{tabular}
\end{center}
\caption{Average $\text{rMSE}_{\mu_1}$ and $\text{rMSE}_{\mu_2}$ for the simulation study where both the location and scale parameters are spatially varying. Results are presented for various choices of the percentage of observed $Z_2$ realizations.} 
\label{Tab1:SimulationStudy_LocScale_Asym_Mu}
\end{table}

\begin{table}[H]
\singlespacing
\begin{center} 
\begin{tabular}{ccccccccc} \toprule 
&\multicolumn{2}{c}{Joint Asym} && \multicolumn{2}{c}{Joint Sym}&& \multicolumn{2}{c}{Ind Sym}\\
\cmidrule{2-3}\cmidrule{5-6}\cmidrule{8-9}
\makecell{Percentage\\Observed}&$\text{rMSE}_{\sigma_1}$ & $\text{rMSE}_{\sigma_2}$ && $\text{rMSE}_{\sigma_1}$&$\text{rMSE}_{\sigma_2}$&& $\text{rMSE}_{\sigma_1}$&$\text{rMSE}_{\sigma_2}$\\
\midrule
1/10& 0.233&0.361&& 0.227& 0.515&& 0.230& 0.567\\
1/15& 0.228&0.351&& 0.218& 0.553&& 0.220& 0.603\\
1/25& 0.237&0.325&& 0.231& 0.570&& 0.230& 0.661\\
1/35& 0.229&0.328&& 0.223& 0.573&& 0.223& 0.712\\
1/50& 0.227&0.302&& 0.218& 0.541&& 0.217& 0.721\\
\bottomrule \end{tabular}
\end{center}
\caption{Average $\text{rMSE}_{\sigma_1}$ and $\text{rMSE}_{\sigma_2}$ for the simulation study where both the location and scale parameters are spatially varying. Results are presented for various choices of the percentage of observed $Z_2$ realizations.} 
\label{Tab1:SimulationStudy_LocScale_Asym_Sigma}
\end{table}

\begin{table}[H]
\singlespacing
\begin{center} 
\begin{tabular}{ccccccccc} \toprule 
&\multicolumn{2}{c}{Joint Asym} && \multicolumn{2}{c}{Joint Sym}&& \multicolumn{2}{c}{Ind Sym}\\
\cmidrule{2-3}\cmidrule{5-6}\cmidrule{8-9}
\makecell{Percentage\\Observed}&$\text{CRPS}_{1}$ & $\text{CRPS}_{2}$ && $\text{CRPS}_{1}$&$\text{CRPS}_{2}$&& $\text{CRPS}_{1}$&$\text{CRPS}_{2}$\\
\midrule
1/10& 0.749&0.943&& 0.750& 1.022&& 0.749& 1.017\\
1/15& 0.749&0.887&& 0.747& 1.002&& 0.749& 0.994\\
1/25& 0.763&0.950&& 0.762& 1.073&& 0.762& 1.078\\
1/35& 0.766&1.008&& 0.765& 1.113&& 0.766& 1.148\\
1/50& 0.755&1.120&& 0.753& 1.201&& 0.753& 1.223\\
\bottomrule \end{tabular}
\end{center}
\caption{Average $\text{CRPS}_1$ and $\text{CRPS}_2$ for the simulation study where both the location and scale parameters are spatially varying. Results are presented for various choices of the percentage of observed $Z_2$ realizations.}
\label{Tab1:SimulationStudy_LocScale_Asym_CRPS}
\vspace*{3\baselineskip}
\end{table}

\begin{table}[H]
\singlespacing
\begin{center} 
\begin{tabular}{ccccccccc} \toprule 
&\multicolumn{2}{c}{Joint Asym} && \multicolumn{2}{c}{Joint Sym}&& \multicolumn{2}{c}{Ind Sym}\\
\cmidrule{2-3}\cmidrule{5-6}\cmidrule{8-9}
\makecell{Percentage\\Observed}&$\text{LogS}_{1}$ & $\text{LogS}_{2}$ && $\text{LogS}_{1}$&$\text{LogS}_{2}$&& $\text{LogS}_{1}$&$\text{LogS}_{2}$\\
\midrule
1/10& 1.604&1.834&& 1.604& 1.900&& 1.604& 1.894\\
1/15& 1.629&1.814&& 1.628& 1.907&& 1.628& 1.883\\
1/25& 1.628&1.912&& 1.628& 1.980&& 1.627& 1.986\\
1/35& 1.643&1.998&& 1.642& 2.042&& 1.642& 2.063\\
1/50& 1.643&2.139&& 1.641& 2.148&& 1.641& 2.147\\
\bottomrule \end{tabular}
\end{center}
\caption{Average $\text{LogS}_1$ and $\text{LogS}_2$ for the simulation study where both the location and scale parameters are spatially varying. Results are presented for various choices of the percentage of observed $Z_2$ realizations.}
\label{Tab1:SimulationStudy_LocScale_Asym_LogS}
\end{table}

\begin{table}[H]
\singlespacing
\begin{center} 
\begin{tabular}{ccccccccc} \toprule 
&\multicolumn{2}{c}{Joint Asym} && \multicolumn{2}{c}{Joint Sym}&& \multicolumn{2}{c}{Ind Sym}\\
\cmidrule{2-3}\cmidrule{5-6}\cmidrule{8-9}
\makecell{Percentage\\Observed}&$\text{rMSE}_1$ & $\text{rMSE}_2$ && $\text{rMSE}_1$&$\text{rMSE}_2$&& $\text{rMSE}_1$&$\text{rMSE}_2$\\
\midrule
1/10& 8.345&11.479&& 7.989& 12.954&& 7.931& 14.861\\
1/15& 8.516&11.313&& 8.266& 13.864&& 8.150& 15.254\\
1/25& 8.507&12.749&& 8.237& 15.741&& 8.237& 18.172\\
1/35& 8.549&13.598&& 8.207& 15.553&& 8.195& 19.977\\
1/50& 8.433&13.742&& 7.997& 16.861&& 8.070& 22.819\\
\bottomrule \end{tabular}
\end{center}
\caption{Average $\text{rMSE}_{RL_{100,1}}$ and $\text{rMSE}_{RL_{100,2}}$ for the simulation study with spatially varying location and scale parameters. Results are presented for various choices of the percentage of observed $Z_2$ realizations.} 
\label{Tab1:SimulationStudy_LocScale_RL100}
\vspace*{3\baselineskip}
\end{table}

\begin{table}[H]
\singlespacing
\begin{center} 
\begin{tabular}{ccccccccc} \toprule 
&\multicolumn{2}{c}{Joint Asym} && \multicolumn{2}{c}{Joint Sym}&& \multicolumn{2}{c}{Ind Sym}\\
\cmidrule{2-3}\cmidrule{5-6}\cmidrule{8-9}
\makecell{Percentage\\Observed}&$\text{CRPS}_{\text{tail},1}$ & $\text{CRPS}_{\text{tail},2}$ && $\text{CRPS}_{\text{tail},1}$&$\text{CRPS}_{\text{tail},2}$&& $\text{CRPS}_{\text{tail},1}$&$\text{CRPS}_{\text{tail},2}$\\
\midrule
1/10& 13.064&14.441&& 13.021& 14.413&& 13.025& 14.554\\
1/15& 13.235&13.803&& 13.236& 13.668&& 13.220& 13.867\\
1/25& 13.446&14.268&& 13.439& 14.048&& 13.454& 14.287\\
1/35& 13.318&14.558&& 13.302& 14.341&& 13.320& 14.606\\
1/50& 13.079&15.089&& 13.062& 14.764&& 13.091& 15.250\\
\bottomrule \end{tabular}
\end{center}
\caption{Average $\text{CRPS}_{\text{tail},1}$ and $\text{CRPS}_{\text{tail},2}$ for the simulation study with spatially varying location and scale parameters. Results are presented for various choices of the percentage of observed $Z_2$ realizations.} 
\label{Tab1:SimulationStudy_LocScale_CRPStail}
\end{table}

\begin{table}[H]
\singlespacing
\begin{center} 
\begin{tabular}{ccccccccc} \toprule 
&\multicolumn{2}{c}{Joint Asym} && \multicolumn{2}{c}{Joint Sym}&& \multicolumn{2}{c}{Ind Sym}\\
\cmidrule{2-3}\cmidrule{5-6}\cmidrule{8-9}
\makecell{Percentage\\Observed}&$\text{LogS}_{\text{tail},1}$ & $\text{LogS}_{\text{tail},2}$ && $\text{LogS}_{\text{tail},1}$&$\text{LogS}_{\text{tail},2}$&& $\text{LogS}_{\text{tail},1}$&$\text{LogS}_{\text{tail},2}$\\
\midrule
1/10& 5.666&5.608&& 5.659& 5.615&& 5.652& 5.615\\
1/15& 5.678&5.552&& 5.681& 5.556&& 5.677& 5.556\\
1/25& 5.689&5.578&& 5.686& 5.550&& 5.688& 5.550\\
1/35& 5.687&5.624&& 5.681& 5.596&& 5.684& 5.596\\
1/50& 5.689&5.704&& 5.686& 5.700&& 5.684& 5.700\\
\bottomrule \end{tabular}
\end{center}
\caption{Average $\text{LogS}_{\text{tail},1}$ and $\text{LogS}_{\text{tail},2}$ for the simulation study with spatially varying location and scale parameters. Results are presented for various choices of the percentage of observed $Z_2$ realizations.} 
\label{Tab1:SimulationStudy_LocScale_LogStail}
\vspace*{3\baselineskip}
\end{table}

\begin{table}[H]
\singlespacing
\begin{center} 
\begin{tabular}{ccccccccc} \toprule 
&\multicolumn{2}{c}{Joint Asym} && \multicolumn{2}{c}{Joint Sym}&& \multicolumn{2}{c}{Ind Sym}\\
\cmidrule{2-3}\cmidrule{5-6}\cmidrule{8-9}
\makecell{Percentage\\Observed}&$\text{rMSE}_1$ & $\text{rMSE}_2$ && $\text{rMSE}_1$&$\text{rMSE}_2$&& $\text{rMSE}_1$&$\text{rMSE}_2$\\
\midrule
1/10& 57.818&65.440&& 57.791& 65.503&& 57.773& 65.622\\
1/15& 62.346&60.375&& 62.357& 60.418&& 62.328& 60.547\\
1/25& 64.534&59.992&& 64.524& 60.005&& 64.537& 60.184\\
1/35& 59.118&95.893&& 59.099& 95.920&& 59.109& 96.080\\
1/50& 62.085&68.398&& 62.052& 68.382&& 62.060& 68.656\\
\bottomrule \end{tabular}
\end{center}
\caption{Mean $\text{rMSE}_{\text{tail},1}$ and $\text{rMSE}_{\text{tail},2}$ for the simulation study with spatially varying location and scale parameters. Results are presented for various choices of the percentage of observed $Z_2$ realizations.} 
\label{Tab1:SimulationStudy_LocScale_rmsetail}
\end{table}

\newpage
\noindent\textbf{Spatially Varying Location Parameter}

Table \ref{Tab1:SimulationStudy_Loc_Asym_Mu} displays the prediction results for the location surfaces for each process. Table \ref{Tab1:SimulationStudy_Loc_Asym_RL10} presents the prediction results for the 10-year return levels. Tables \ref{Tab1:SimulationStudy_Loc_Asym_CRPS} and \ref{Tab1:SimulationStudy_Loc_Asym_LogS} show the average CRPS and LogS for each process. For each metric, the joint asymmetric model consistently yields more accurate predictions than the joint and independent symmetric models for the \(Z_2\) process, with similar results for the \(Z_1\) process.

\begin{table}[H]
\singlespacing
\begin{center} 
\begin{tabular}{ccccccccc} \toprule 
&\multicolumn{2}{c}{Joint Asym} && \multicolumn{2}{c}{Joint Sym}&& \multicolumn{2}{c}{Ind Sym}\\
\cmidrule{2-3}\cmidrule{5-6}\cmidrule{8-9}
\makecell{Percentage\\Observed}&$\text{rMSE}_{\mu_1}$ & $\text{rMSE}_{\mu_2}$ && $\text{rMSE}_{\mu_1}$&$\text{rMSE}_{\mu_2}$&& $\text{rMSE}_{\mu_1}$&$\text{rMSE}_{\mu_2}$\\
\midrule
1/10& 0.155&0.356&& 0.142& 0.537&& 0.142& 0.506\\
1/15& 0.154&0.419&& 0.142& 0.569&& 0.142& 0.580\\
1/25& 0.155&0.558&& 0.141& 0.689&& 0.142& 0.679\\
1/35& 0.154&0.666&& 0.141& 0.763&& 0.141& 0.778\\
1/50& 0.155&0.801&& 0.144& 0.924&& 0.143& 0.946\\
\bottomrule \end{tabular}
\end{center}
\caption{Average $\text{rMSE}_{\mu_1}$ and $\text{rMSE}_{\mu_2}$ for the simulation study where only the location parameter is spatially varying. Results are presented for various choices of the percentage of observed $Z_2$ realizations.} 
\label{Tab1:SimulationStudy_Loc_Asym_Mu}
\end{table}

\begin{table}[H]
\vspace*{3\baselineskip}
\singlespacing
\begin{center} 
\begin{tabular}{ccccccccc} \toprule 
&\multicolumn{2}{c}{Joint Asym} && \multicolumn{2}{c}{Joint Sym}&& \multicolumn{2}{c}{Ind Sym}\\
\cmidrule{2-3}\cmidrule{5-6}\cmidrule{8-9}
\makecell{Percentage\\Observed}&$\text{rMSE}_1$ & $\text{rMSE}_2$ && $\text{rMSE}_1$&$\text{rMSE}_2$&& $\text{rMSE}_1$&$\text{rMSE}_2$\\
\midrule
1/10& 0.162&0.388&& 0.149& 0.595&& 0.149& 0.566\\
1/15& 0.161&0.461&& 0.150& 0.643&& 0.150& 0.662\\
1/25& 0.155&0.801&& 0.144& 0.924&& 0.143& 0.946\\
1/35& 0.163&0.723&& 0.151& 0.852&& 0.151& 0.869\\
1/50& 0.163&0.858&& 0.153& 1.038&& 0.152& 1.086\\
\bottomrule \end{tabular}
\end{center}
\caption{Average $\text{rMSE}_{RL_{10,1}}$ and $\text{rMSE}_{RL_{10,2}}$ for the simulation study where only the location parameter is spatially varying. Results are presented for various choices of the percentage of observed $Z_2$ realizations.} 
\label{Tab1:SimulationStudy_Loc_Asym_RL10}
\end{table}

\begin{table}[H]
\singlespacing
\begin{center} 
\begin{tabular}{ccccccccc} \toprule 
&\multicolumn{2}{c}{Joint Asym} && \multicolumn{2}{c}{Joint Sym}&& \multicolumn{2}{c}{Ind Sym}\\
\cmidrule{2-3}\cmidrule{5-6}\cmidrule{8-9}
\makecell{Percentage\\Observed}&$\text{CRPS}_{1}$ & $\text{CRPS}_{2}$ && $\text{CRPS}_{1}$&$\text{CRPS}_{2}$&& $\text{CRPS}_{1}$&$\text{CRPS}_{2}$\\
\midrule
1/10& 0.615&0.629&& 0.613& 0.676&& 0.613& 0.670\\
1/15& 0.614&0.639&& 0.613& 0.682&& 0.613& 0.685\\
1/25& 0.614&0.656&& 0.613& 0.703&& 0.613& 0.698\\
1/35& 0.613&0.669&& 0.612& 0.706&& 0.612& 0.712\\
1/50& 0.613&0.696&& 0.612& 0.747&& 0.612& 0.757\\
\bottomrule \end{tabular}
\end{center}
\caption{Average $\text{CRPS}_1$ and $\text{CRPS}_2$ for the simulation study where only the location parameter is spatially varying. Results are presented for various choices of the percentage of observed $Z_2$ realizations.} 
\label{Tab1:SimulationStudy_Loc_Asym_CRPS}
\end{table}

\begin{table}[H]
\vspace*{3\baselineskip}
\singlespacing
\begin{center} 
\begin{tabular}{ccccccccc} \toprule 
&\multicolumn{2}{c}{Joint Asym} && \multicolumn{2}{c}{Joint Sym}&& \multicolumn{2}{c}{Ind Sym}\\
\cmidrule{2-3}\cmidrule{5-6}\cmidrule{8-9}
\makecell{Percentage\\Observed}&$\text{LogS}_{1}$ & $\text{LogS}_{2}$ && $\text{LogS}_{1}$&$\text{LogS}_{2}$&& $\text{LogS}_{1}$&$\text{LogS}_{2}$\\
\midrule
1/10& 1.393&1.455&& 1.390& 1.559&& 1.391& 1.543\\
1/15& 1.393&1.480&& 1.390& 1.576&& 1.390& 1.574\\
1/25& 1.393&1.532&& 1.390& 1.619&& 1.390& 1.615\\
1/35& 1.392&1.579&& 1.389& 1.653&& 1.389& 1.662\\
1/50& 1.392&1.640&& 1.390& 1.729&& 1.389& 1.737\\
\bottomrule \end{tabular}
\end{center}
\caption{Average $\text{LogS}_1$ and $\text{LogS}_2$ for the simulation study where only the location parameter is spatially varying. Results are presented for various choices of the percentage of observed $Z_2$ realizations.} 
\label{Tab1:SimulationStudy_Loc_Asym_LogS}
\end{table}

\begin{table}[H]
\singlespacing
\begin{center} 
\begin{tabular}{ccccccccc} \toprule 
&\multicolumn{2}{c}{Joint Asym} && \multicolumn{2}{c}{Joint Sym}&& \multicolumn{2}{c}{Ind Sym}\\
\cmidrule{2-3}\cmidrule{5-6}\cmidrule{8-9}
\makecell{Percentage\\Observed}&$\text{rMSE}_1$ & $\text{rMSE}_2$ && $\text{rMSE}_1$&$\text{rMSE}_2$&& $\text{rMSE}_1$&$\text{rMSE}_2$\\
\midrule
1/10& 1.154&1.713&& 1.022& 3.128&& 1.025& 3.007\\
1/15& 1.156&1.670&& 1.068& 2.966&& 1.076& 2.967\\
1/25& 1.212&1.875&& 1.081& 2.769&& 1.109& 2.792\\
1/35& 1.226&2.089&& 1.119& 2.867&& 1.132& 3.028\\
1/50& 1.252&2.357&& 1.152& 2.825&& 1.134& 2.774\\
\bottomrule \end{tabular}
\end{center}
\caption{Average $\text{rMSE}_{RL_{100,1}}$ and $\text{rMSE}_{RL_{100,2}}$ for the simulation study with spatially varying location parameter. Results are presented for various choices of the percentage of observed $Z_2$ realizations.} 
\label{Tab1:SimulationStudy_LocScale_RL100}
\vspace*{3\baselineskip}
\end{table}

\begin{table}[H]
\singlespacing
\begin{center} 
\begin{tabular}{ccccccccc} \toprule 
&\multicolumn{2}{c}{Joint Asym} && \multicolumn{2}{c}{Joint Sym}&& \multicolumn{2}{c}{Ind Sym}\\
\cmidrule{2-3}\cmidrule{5-6}\cmidrule{8-9}
\makecell{Percentage\\Observed}&$\text{CRPS}_{\text{tail},1}$ & $\text{CRPS}_{\text{tail},2}$ && $\text{CRPS}_{\text{tail},1}$&$\text{CRPS}_{\text{tail},2}$&& $\text{CRPS}_{\text{tail},1}$&$\text{CRPS}_{\text{tail},2}$\\
\midrule
1/10& 8.258&8.118&& 8.261& 8.076&& 8.257& 8.063\\
1/15& 8.214&8.105&& 8.213& 8.064&& 8.214& 8.062\\
1/25& 8.220&8.094&& 8.225& 8.082&& 8.220& 8.052\\
1/35& 8.200&8.138&& 8.198& 8.107&& 8.198& 8.081\\
1/50& 8.208&8.195&& 8.210& 8.109&& 8.213& 8.097\\
\bottomrule \end{tabular}
\end{center}
\caption{Average $\text{CRPS}_{\text{tail},1}$ and $\text{CRPS}_{\text{tail},2}$ for the simulation study with spatially varying location parameter. Results are presented for various choices of the percentage of observed $Z_2$ realizations.} 
\label{Tab1:SimulationStudy_LocScale_CRPStail}
\end{table}

\begin{table}[H]
\singlespacing
\begin{center} 
\begin{tabular}{ccccccccc} \toprule 
&\multicolumn{2}{c}{Joint Asym} && \multicolumn{2}{c}{Joint Sym}&& \multicolumn{2}{c}{Ind Sym}\\
\cmidrule{2-3}\cmidrule{5-6}\cmidrule{8-9}
\makecell{Percentage\\Observed}&$\text{LogS}_{\text{tail},1}$ & $\text{LogS}_{\text{tail},2}$ && $\text{LogS}_{\text{tail},1}$&$\text{LogS}_{\text{tail},2}$&& $\text{LogS}_{\text{tail},1}$&$\text{LogS}_{\text{tail},2}$\\
\midrule
1/10& 5.569&5.532&& 5.566& 5.496&& 5.565& 5.496\\
1/15& 5.555&5.521&& 5.556& 5.484&& 5.554& 5.484\\
1/25& 5.561&5.508&& 5.556& 5.468&& 5.558& 5.469\\
1/35& 5.552&5.539&& 5.555& 5.484&& 5.554& 5.484\\
1/50& 5.559&5.536&& 5.557& 5.454&& 5.557& 5.454\\
\bottomrule \end{tabular}
\end{center}
\caption{Average $\text{LogS}_{\text{tail},1}$ and $\text{LogS}_{\text{tail},2}$ for the simulation study with spatially varying location parameter. Results are presented for various choices of the percentage of observed $Z_2$ realizations.} 
\label{Tab1:SimulationStudy_LocScale_LogStail}
\vspace*{3\baselineskip}
\end{table}

\begin{table}[H]
\singlespacing
\begin{center} 
\begin{tabular}{ccccccccc} \toprule 
&\multicolumn{2}{c}{Joint Asym} && \multicolumn{2}{c}{Joint Sym}&& \multicolumn{2}{c}{Ind Sym}\\
\cmidrule{2-3}\cmidrule{5-6}\cmidrule{8-9}
\makecell{Percentage\\Observed}&$\text{rMSE}_1$ & $\text{rMSE}_2$ && $\text{rMSE}_1$&$\text{rMSE}_2$&& $\text{rMSE}_1$&$\text{rMSE}_2$\\
\midrule
1/10& 29.649&26.823&& 29.646& 26.877&& 29.646& 26.871\\
1/15& 29.273&29.396&& 29.270& 29.452&& 29.271& 29.452\\
1/25& 28.307&35.688&& 28.291& 35.897&& 28.305& 35.749\\
1/35& 28.827&32.294&& 28.824& 32.369&& 28.824& 32.366\\
1/50& 27.785&25.072&& 27.782& 25.146&& 27.782& 25.146\\
\bottomrule \end{tabular}
\end{center}
\caption{Mean $\text{rMSE}_{\text{tail},1}$ and $\text{rMSE}_{\text{tail},2}$ for the simulation study with spatially varying location parameter. Results are presented for various choices of the percentage of observed $Z_2$ realizations.} 
\label{Tab1:SimulationStudy_LocScale_rmsetail}
\end{table}

\newpage
\subsection{Symmetric Data Generation}

We next study the simulation results in a scenario where both the location and scale parameters are spatially varying, while the shape parameters remain fixed but unknown. Then we study the case where only the location parameter is spatially varying. We focus on the case where the latent spatial surfaces for each process are negatively correlated with a symmetric cross-covariance structure.

\noindent\textbf{Spatially Varying Location and Scale Parameters}

Tables \ref{Tab1:SimulationStudy_LocScale_Sym_Mu} and \ref{Tab1:SimulationStudy_LocScale_Sym_Sigma} display the prediction results for the location and scale surfaces, respectively. Table \ref{Tab1:SimulationStudy_LocScale_Sym_RL10} presents the prediction results for the 10-year return levels. Tables \ref{Tab1:SimulationStudy_LocScale_Sym_CRPS} and \ref{Tab1:SimulationStudy_LocScale_Sym_LogS} show the average CRPS and LogS for each process. The joint symmetric model outperforms the joint asymmetric model for the $Z_2$ process, especially when \(Z_2\) observations are sparse. However, the difference in performance is relatively small and the joint symmetric model is correctly specified. Notably, the joint asymmetric model still provides more accurate predictions than the independent model. 

\begin{table}[H]
\singlespacing
\begin{center} 
\begin{tabular}{ccccccccc} \toprule 
&\multicolumn{2}{c}{Joint Asym} && \multicolumn{2}{c}{Joint Sym}&& \multicolumn{2}{c}{Ind Sym}\\
\cmidrule{2-3}\cmidrule{5-6}\cmidrule{8-9}
\makecell{Percentage\\Observed}&$\text{rMSE}_{\mu_1}$ & $\text{rMSE}_{\mu_2}$ && $\text{rMSE}_{\mu_1}$&$\text{rMSE}_{\mu_2}$&& $\text{rMSE}_{\mu_1}$&$\text{rMSE}_{\mu_2}$\\
\midrule
1/10& 0.258&0.305&& 0.228& 0.233&& 0.246& 0.468\\
1/15& 0.251&0.317&& 0.232& 0.241&& 0.250& 0.515\\
1/25& 0.264&0.346&& 0.232& 0.261&& 0.250& 0.542\\
1/35& 0.259&0.393&& 0.254& 0.287&& 0.253& 0.575\\
1/50& 0.260&0.448&& 0.250& 0.303&& 0.251& 0.611\\
\bottomrule \end{tabular}
\end{center}
\caption{Average $\text{rMSE}_{\mu_{1}}$ and $\text{rMSE}_{\mu_{2}}$ for the simulation study where both the location
and scale parameters are spatially varying. Results are presented for various choices of the percentage of observed $Z_2$ realizations.} 
\label{Tab1:SimulationStudy_LocScale_Sym_Mu}
\end{table}

\begin{table}[H]
\vspace*{3\baselineskip}
\singlespacing
\begin{center} 
\begin{tabular}{ccccccccc} \toprule 
&\multicolumn{2}{c}{Joint Asym} && \multicolumn{2}{c}{Joint Sym}&& \multicolumn{2}{c}{Ind Sym}\\
\cmidrule{2-3}\cmidrule{5-6}\cmidrule{8-9}
\makecell{Percentage\\Observed}&$\text{rMSE}_{\sigma_1}$ & $\text{rMSE}_{\sigma_2}$ && $\text{rMSE}_{\sigma_1}$&$\text{rMSE}_{\sigma_2}$&& $\text{rMSE}_{\sigma_1}$&$\text{rMSE}_{\sigma_2}$\\
\midrule
1/10& 0.226&0.257&& 0.216& 0.265&& 0.218& 0.432\\
1/15& 0.235&0.257&& 0.227& 0.277&& 0.227& 0.462\\
1/25& 0.220&0.243&& 0.213& 0.260&& 0.213& 0.439\\
1/35& 0.227&0.259&& 0.221& 0.280&& 0.222& 0.522\\
1/50& 0.232&0.275&& 0.222& 0.289&& 0.225& 0.588\\
\bottomrule \end{tabular}
\end{center}
\caption{Average $\text{rMSE}_{\sigma_{1}}$ and $\text{rMSE}_{\sigma_{2}}$ for the simulation study where both the location and scale parameters are spatially varying. Results are presented for various choices of the percentage of observed $Z_2$ realizations.} 
\label{Tab1:SimulationStudy_LocScale_Sym_Sigma}
\vspace*{3\baselineskip}
\end{table}

\begin{table}[H]
\singlespacing
\begin{center} 
\begin{tabular}{ccccccccc} \toprule 
&\multicolumn{2}{c}{Joint Asym} && \multicolumn{2}{c}{Joint Sym}&& \multicolumn{2}{c}{Ind Sym}\\
\cmidrule{2-3}\cmidrule{5-6}\cmidrule{8-9}
\makecell{Percentage\\Observed}&$\text{rMSE}_1$ & $\text{rMSE}_2$ && $\text{rMSE}_1$&$\text{rMSE}_2$&& $\text{rMSE}_1$&$\text{rMSE}_2$\\
\midrule
1/10& 1.567&1.969&& 1.448& 1.829&& 1.481& 2.638\\
1/15& 1.547&2.083&& 1.449& 1.913&& 1.483& 2.905\\
1/25& 1.621&2.451&& 1.516& 2.197&& 1.538& 3.389\\
1/35& 1.568&2.592&& 1.523& 2.409&& 1.505& 3.730\\
1/50& 1.589&2.478&& 1.489& 2.260&& 1.504& 3.744\\
\bottomrule \end{tabular}
\end{center}
\caption{Average $\text{rMSE}_{RL_{10,1}}$ and $\text{rMSE}_{RL_{10,2}}$ for the simulation study where both the location and scale parameters are spatially varying. Results are presented for various choices of the percentage of observed $Z_2$ realizations.} 
\label{Tab1:SimulationStudy_LocScale_Sym_RL10}
\end{table}

\begin{table}[H]
\singlespacing
\begin{center} 
\begin{tabular}{ccccccccc} \toprule 
&\multicolumn{2}{c}{Joint Asym} && \multicolumn{2}{c}{Joint Sym}&& \multicolumn{2}{c}{Ind Sym}\\
\cmidrule{2-3}\cmidrule{5-6}\cmidrule{8-9}
\makecell{Percentage\\Observed}&$\text{CRPS}_{1}$ & $\text{CRPS}_{2}$ && $\text{CRPS}_{1}$&$\text{CRPS}_{2}$&& $\text{CRPS}_{1}$&$\text{CRPS}_{2}$\\
\midrule
1/10& 0.758&0.770&& 0.756& 0.763&& 0.757& 0.815\\
1/15& 0.708&0.786&& 0.705& 0.779&& 0.708& 0.847\\
1/25& 0.763&0.925&& 0.763& 0.913&& 0.762& 0.979\\
1/35& 0.749&0.818&& 0.748& 0.806&& 0.749& 0.872\\
1/50& 0.769&0.737&& 0.768& 0.709&& 0.768& 0.810\\
\bottomrule \end{tabular}
\end{center}
\caption{Average $\text{CRPS}_{1}$ and $\text{CRPS}_{2}$ for the simulation study where both the location and scale parameters are spatially varying. Results are presented for various choices of the percentage of observed $Z_2$ realizations.} 
\label{Tab1:SimulationStudy_LocScale_Sym_CRPS}
\vspace*{3\baselineskip}
\end{table}

\begin{table}[H]
\singlespacing
\begin{center} 
\begin{tabular}{ccccccccc} \toprule 
&\multicolumn{2}{c}{Joint Asym} && \multicolumn{2}{c}{Joint Sym}&& \multicolumn{2}{c}{Ind Sym}\\
\cmidrule{2-3}\cmidrule{5-6}\cmidrule{8-9}
\makecell{Percentage\\Observed}&$\text{LogS}_{1}$ & $\text{LogS}_{2}$ && $\text{LogS}_{1}$&$\text{LogS}_{2}$&& $\text{LogS}_{1}$&$\text{LogS}_{2}$\\
\midrule
1/10& 1.625&1.583&& 1.622& 1.556&& 1.624& 1.637\\
1/15& 1.566&1.632&& 1.565& 1.603&& 1.567& 1.695\\
1/25& 1.640&1.809&& 1.639& 1.777&& 1.639& 1.854\\
1/35& 1.642&1.737&& 1.641& 1.679&& 1.642& 1.775\\
1/50& 1.647&1.671&& 1.645& 1.567&& 1.645& 1.729\\
\bottomrule \end{tabular}
\end{center}
\caption{Average $\text{LogS}_{1}$ and $\text{LogS}_{2}$ for the simulation study where both the location and scale parameters are spatially varying. Results are presented for various choices of the percentage of observed $Z_2$ realizations.}
\label{Tab1:SimulationStudy_LocScale_Sym_LogS}
\end{table}

\begin{table}[H]
\singlespacing
\begin{center} 
\begin{tabular}{ccccccccc} \toprule 
&\multicolumn{2}{c}{Joint Asym} && \multicolumn{2}{c}{Joint Sym}&& \multicolumn{2}{c}{Ind Sym}\\
\cmidrule{2-3}\cmidrule{5-6}\cmidrule{8-9}
\makecell{Percentage\\Observed}&$\text{rMSE}_1$ & $\text{rMSE}_2$ && $\text{rMSE}_1$&$\text{rMSE}_2$&& $\text{rMSE}_1$&$\text{rMSE}_2$\\
\midrule
1/10& 7.953&8.743&& 7.594& 8.755&& 7.546& 11.407\\
1/15& 7.986&9.229&& 7.664& 9.193&& 7.621& 12.161\\
1/25& 8.298&10.774&& 7.907& 10.283&& 7.848& 14.156\\
1/35& 8.003&11.362&& 7.699& 10.996&& 7.702& 15.699\\
1/50& 8.081&10.720&& 7.650& 10.412&& 7.738& 15.655\\
\bottomrule \end{tabular}
\end{center}
\caption{Average $\text{rMSE}_{RL_{100,1}}$ and $\text{rMSE}_{RL_{100,2}}$ for the simulation study with spatially varying location and scale parameters. Results are presented for various choices of the percentage of observed $Z_2$ realizations.} 
\label{Tab1:SimulationStudy_RL100}
\vspace*{3\baselineskip}
\end{table}

\begin{table}[H]
\singlespacing
\begin{center} 
\begin{tabular}{ccccccccc} \toprule 
&\multicolumn{2}{c}{Joint Asym} && \multicolumn{2}{c}{Joint Sym}&& \multicolumn{2}{c}{Ind Sym}\\
\cmidrule{2-3}\cmidrule{5-6}\cmidrule{8-9}
\makecell{Percentage\\Observed}&$\text{CRPS}_{\text{tail},1}$ & $\text{CRPS}_{\text{tail},2}$ && $\text{CRPS}_{\text{tail},1}$&$\text{CRPS}_{\text{tail},2}$&& $\text{CRPS}_{\text{tail},1}$&$\text{CRPS}_{\text{tail},2}$\\
\midrule
1/10& 12.552&12.478&& 12.548& 12.504&& 12.546& 12.528\\
1/15& 12.193&13.007&& 12.199& 13.067&& 12.163& 13.063\\
1/25& 13.165&14.627&& 13.152& 14.685&& 13.138& 14.791\\
1/35& 13.003&13.323&& 13.001& 13.361&& 13.007& 13.642\\
1/50& 12.896&11.673&& 12.881& 11.716&& 12.881& 11.898\\
\bottomrule \end{tabular}
\end{center}
\caption{Average $\text{CRPS}_{\text{tail},1}$ and $\text{CRPS}_{\text{tail},2}$ for the simulation study with spatially varying location and scale parameters. Results are presented for various choices of the percentage of observed $Z_2$ realizations.} 
\label{Tab1:SimulationStudy_CRPStail}
\end{table}

\begin{table}[H] 
\singlespacing
\begin{center} 
\begin{tabular}{ccccccccc} \toprule 
&\multicolumn{2}{c}{Joint Asym} && \multicolumn{2}{c}{Joint Sym}&& \multicolumn{2}{c}{Ind Sym}\\
\cmidrule{2-3}\cmidrule{5-6}\cmidrule{8-9}
\makecell{Percentage\\Observed}&$\text{LogS}_{\text{tail},1}$ & $\text{LogS}_{\text{tail},2}$ && $\text{LogS}_{\text{tail},1}$&$\text{LogS}_{\text{tail},2}$&& $\text{LogS}_{\text{tail},1}$&$\text{LogS}_{\text{tail},2}$\\
\midrule
1/10& 5.640&5.525&& 5.642& 5.544&& 5.637& 5.544\\
1/15& 5.602&5.582&& 5.601& 5.588&& 5.595& 5.588\\
1/25& 5.712&5.737&& 5.708& 5.777&& 5.701& 5.777\\
1/35& 5.667&5.651&& 5.662& 5.736&& 5.664& 5.736\\
1/50& 5.665&5.469&& 5.661& 5.536&& 5.663& 5.536\\
\bottomrule \end{tabular}
\end{center}
\caption{Average $\text{LogS}_{\text{tail},1}$ and $\text{LogS}_{\text{tail},2}$ for the simulation study with spatially varying location and scale parameters. Results are presented for various choices of the percentage of observed $Z_2$ realizations.}
\label{Tab1:SimulationStudy_LogStail}
\vspace*{3\baselineskip}
\end{table}

\begin{table}[H]
\singlespacing
\begin{center} 
\begin{tabular}{ccccccccc} \toprule 
&\multicolumn{2}{c}{Joint Asym} && \multicolumn{2}{c}{Joint Sym}&& \multicolumn{2}{c}{Ind Sym}\\
\cmidrule{2-3}\cmidrule{5-6}\cmidrule{8-9}
\makecell{Percentage\\Observed}&$\text{rMSE}_1$ & $\text{rMSE}_2$ && $\text{rMSE}_1$&$\text{rMSE}_2$&& $\text{rMSE}_1$&$\text{rMSE}_2$\\
\midrule
1/10& 55.412&52.256&& 55.408& 52.313&& 55.395& 52.419\\
1/15& 53.150&53.129&& 53.143& 53.222&& 53.112& 53.282\\
1/25& 56.774&62.402&& 56.757& 62.510&& 56.741& 62.607\\
1/35& 55.738&100.198&& 55.718& 100.290&& 55.721& 100.515\\
1/50& 60.517&48.053&& 60.499& 48.215&& 60.504& 48.353\\
\bottomrule \end{tabular}
\end{center}
\caption{Mean $\text{rMSE}_{\text{tail},1}$ and $\text{rMSE}_{\text{tail},2}$ for the simulation study with spatially varying location and scale parameters. Results are presented for various choices of the percentage of observed $Z_2$ realizations.} 
\label{Tab1:SimulationStudy_rmsetail}
\end{table}

\newpage
\noindent\textbf{Spatially Varying Location Parameter}

Table \ref{Tab1:SimulationStudy_Loc_Sym_Mu} displays the prediction results for the location surfaces for each process. Table \ref{Tab1:SimulationStudy_Loc_Sym_RL10} presents the prediction results for the 10-year return levels. For each metric, the joint symmetric model yields more accurate predictions than the joint asymmetric model for the \(Z_2\) process, with similar results for the \(Z_1\) process. The difference in performance is more pronounced for the \(Z_2\) process when the proportion of missing \(Z_2\) observations is larger. However, the difference in performance is relatively small and the joint symmetric model is correctly specified. Notably, the joint asymmetric
model still provides more accurate predictions than the independent model. Tables \ref{Tab1:SimulationStudy_Loc_Sym_CRPS} and \ref{Tab1:SimulationStudy_Loc_Sym_LogS} show the average CRPS and LogS for each process. The joint asymmetric model performs similarly to the joint symmetric model for both the $Z_1$ and \(Z_2\) processes, despite the joint asymmetric model being misspecified. Notably, both the joint asymmetric and joint symmetric models outperform the independent model.

\begin{table}[H]
\vspace*{3\baselineskip}
\singlespacing
\begin{center} 
\begin{tabular}{ccccccccc} \toprule 
&\multicolumn{2}{c}{Joint Asym} && \multicolumn{2}{c}{Joint Sym}&& \multicolumn{2}{c}{Ind Sym}\\
\cmidrule{2-3}\cmidrule{5-6}\cmidrule{8-9}
\makecell{Percentage\\Observed}&$\text{rMSE}_{\mu_1}$ & $\text{rMSE}_{\mu_2}$ && $\text{rMSE}_{\mu_1}$&$\text{rMSE}_{\mu_2}$&& $\text{rMSE}_{\mu_1}$&$\text{rMSE}_{\mu_2}$\\
\midrule
1/10& 0.150&0.159&& 0.141& 0.148&& 0.141& 0.269\\
1/15& 0.153&0.168&& 0.143& 0.155&& 0.144& 0.311\\
1/25& 0.153&0.176&& 0.142& 0.159&& 0.143& 0.372\\
1/35& 0.151&0.185&& 0.137& 0.166&& 0.139& 0.405\\
1/50& 0.152&0.196&& 0.141& 0.179&& 0.141& 0.482\\
\bottomrule \end{tabular}
\end{center}
\caption{Average $\text{rMSE}_{\mu_{1}}$ and $\text{rMSE}_{\mu_{2}}$ for the simulation study with spatially varying location parameter. Results are presented for various choices of the percentage of observed $Z_2$ realizations.} 
\label{Tab1:SimulationStudy_Loc_Sym_Mu}
\end{table}

\begin{table}[H] 
\singlespacing
\begin{center} 
\begin{tabular}{ccccccccc} \toprule 
&\multicolumn{2}{c}{Joint Asym} && \multicolumn{2}{c}{Joint Sym}&& \multicolumn{2}{c}{Ind Sym}\\
\cmidrule{2-3}\cmidrule{5-6}\cmidrule{8-9}
\makecell{Percentage\\Observed}&$\text{rMSE}_1$ & $\text{rMSE}_2$ && $\text{rMSE}_1$&$\text{rMSE}_2$&& $\text{rMSE}_1$&$\text{rMSE}_2$\\
\midrule
1/10& 0.157&0.209&& 0.148& 0.202&& 0.148& 0.307\\
1/15& 0.160&0.231&& 0.150& 0.221&& 0.151& 0.357\\
1/25& 0.162&0.279&& 0.151& 0.269&& 0.152& 0.441\\
1/35& 0.159&0.308&& 0.146& 0.297&& 0.147& 0.488\\
1/50& 0.161&0.365&& 0.151& 0.354&& 0.151& 0.597\\
\bottomrule \end{tabular}
\end{center}
\caption{Average $\text{rMSE}_{RL_{10,1}}$ and $\text{rMSE}_{RL_{10,2}}$ for the simulation study with spatially varying location parameter. Results are presented for various choices of the percentage of observed $Z_2$ realizations.}
\label{Tab1:SimulationStudy_Loc_Sym_RL10}
\vspace*{3\baselineskip}
\end{table}

\begin{table}[H] 
\singlespacing
\begin{center} 
\begin{tabular}{ccccccccc} \toprule 
&\multicolumn{2}{c}{Joint Asym} && \multicolumn{2}{c}{Joint Sym}&& \multicolumn{2}{c}{Ind Sym}\\
\cmidrule{2-3}\cmidrule{5-6}\cmidrule{8-9}
\makecell{Percentage\\Observed}&$\text{CRPS}_{1}$ & $\text{CRPS}_{2}$ && $\text{CRPS}_{1}$&$\text{CRPS}_{2}$&& $\text{CRPS}_{1}$&$\text{CRPS}_{2}$\\
\midrule
1/10& 0.612&0.613&& 0.612& 0.613&& 0.612& 0.627\\
1/15& 0.614&0.615&& 0.613& 0.615&& 0.613& 0.634\\
1/25& 0.613&0.613&& 0.612& 0.613&& 0.612& 0.637\\
1/35& 0.613&0.613&& 0.612& 0.612&& 0.612& 0.642\\
1/50& 0.612&0.617&& 0.612& 0.617&& 0.611& 0.655\\
\bottomrule \end{tabular}
\end{center}
\caption{Average $\text{CRPS}_{1}$ and $\text{CRPS}_{2}$ for the simulation study with spatially varying location parameter. Results are presented for various choices of the percentage of observed $Z_2$ realizations.}
\label{Tab1:SimulationStudy_Loc_Sym_CRPS}
\end{table}

\begin{table}[H]
\singlespacing
\begin{center} 
\begin{tabular}{ccccccccc} \toprule 
&\multicolumn{2}{c}{Joint Asym} && \multicolumn{2}{c}{Joint Sym}&& \multicolumn{2}{c}{Ind Sym}\\
\cmidrule{2-3}\cmidrule{5-6}\cmidrule{8-9}
\makecell{Percentage\\Observed}&$\text{LogS}_{1}$ & $\text{LogS}_{2}$ && $\text{LogS}_{1}$&$\text{LogS}_{2}$&& $\text{LogS}_{1}$&$\text{LogS}_{2}$\\
\midrule
1/10& 1.391&1.395&& 1.389& 1.394&& 1.389& 1.439\\
1/15& 1.392&1.398&& 1.390& 1.395&& 1.390& 1.458\\
1/25& 1.391&1.398&& 1.388& 1.393&& 1.389& 1.479\\
1/35& 1.391&1.403&& 1.388& 1.400&& 1.389& 1.502\\
1/50& 1.391&1.408&& 1.389& 1.404&& 1.388& 1.532\\
\bottomrule \end{tabular}
\end{center}
\caption{Average $\text{LogS}_{1}$ and $\text{LogS}_{2}$ for the simulation study where only the location parameter is spatially varying. Results are presented for various choices of the percentage of observed $Z_2$ realizations.}
\label{Tab1:SimulationStudy_Loc_Sym_LogS}
\vspace*{3\baselineskip}
\end{table}

\begin{table}[H]
\singlespacing
\begin{center} 
\begin{tabular}{ccccccccc} \toprule 
&\multicolumn{2}{c}{Joint Asym} && \multicolumn{2}{c}{Joint Sym}&& \multicolumn{2}{c}{Ind Sym}\\
\cmidrule{2-3}\cmidrule{5-6}\cmidrule{8-9}
\makecell{Percentage\\Observed}&$\text{rMSE}_1$ & $\text{rMSE}_2$ && $\text{rMSE}_1$&$\text{rMSE}_2$&& $\text{rMSE}_1$&$\text{rMSE}_2$\\
\midrule
1/10& 1.198&1.116&& 1.118& 1.137&& 1.092& 1.863\\
1/15& 1.196&1.323&& 1.106& 1.366&& 1.096& 2.154\\
1/25& 1.229&1.724&& 1.116& 1.785&& 1.111& 2.488\\
1/35& 1.159&1.743&& 1.012& 1.708&& 1.028& 2.280\\
1/50& 1.244&2.203&& 1.155& 2.234&& 1.142& 2.578\\
\bottomrule \end{tabular}
\end{center}
\caption{Average $\text{rMSE}_{RL_{100,1}}$ and $\text{rMSE}_{RL_{100,2}}$ for the simulation study where only the location parameter is spatially varying. Results are presented for various choices of the percentage of observed $Z_2$ realizations.} 
\label{Tab1:SimulationStudy_RL100}
\end{table}

\begin{table}[H]
\singlespacing
\begin{center} 
\begin{tabular}{ccccccccc} \toprule 
&\multicolumn{2}{c}{Joint Asym} && \multicolumn{2}{c}{Joint Sym}&& \multicolumn{2}{c}{Ind Sym}\\
\cmidrule{2-3}\cmidrule{5-6}\cmidrule{8-9}
\makecell{Percentage\\Observed}&$\text{CRPS}_{\text{tail},1}$ & $\text{CRPS}_{\text{tail},2}$ && $\text{CRPS}_{\text{tail},1}$&$\text{CRPS}_{\text{tail},2}$&& $\text{CRPS}_{\text{tail},1}$&$\text{CRPS}_{\text{tail},2}$\\
\midrule
1/10& 8.203&8.221&& 8.205& 8.222&& 8.204& 8.207\\
1/15& 8.218&8.203&& 8.215& 8.200&& 8.215& 8.184\\
1/25& 8.195&8.207&& 8.196& 8.205&& 8.197& 8.188\\
1/35& 8.230&8.190&& 8.229& 8.185&& 8.228& 8.169\\
1/50& 8.213&8.169&& 8.215& 8.167&& 8.216& 8.126\\
\bottomrule \end{tabular}
\end{center}
\caption{Average $\text{CRPS}_{\text{tail},1}$ and $\text{CRPS}_{\text{tail},2}$ for the simulation study where only the location parameter is spatially varying. Results are presented for various choices of the percentage of observed $Z_2$ realizations.} 
\label{Tab1:SimulationStudy_CRPStail}
\vspace*{3\baselineskip}
\end{table}

\begin{table}[H]
\singlespacing
\begin{center} 
\begin{tabular}{ccccccccc} \toprule 
&\multicolumn{2}{c}{Joint Asym} && \multicolumn{2}{c}{Joint Sym}&& \multicolumn{2}{c}{Ind Sym}\\
\cmidrule{2-3}\cmidrule{5-6}\cmidrule{8-9}
\makecell{Percentage\\Observed}&$\text{LogS}_{\text{tail},1}$ & $\text{LogS}_{\text{tail},2}$ && $\text{LogS}_{\text{tail},1}$&$\text{LogS}_{\text{tail},2}$&& $\text{LogS}_{\text{tail},1}$&$\text{LogS}_{\text{tail},2}$\\
\midrule
1/10& 5.558&5.558&& 5.551& 5.559&& 5.555& 5.559\\
1/15& 5.559&5.551&& 5.558& 5.548&& 5.557& 5.548\\
1/25& 5.558&5.569&& 5.554& 5.564&& 5.555& 5.564\\
1/35& 5.563&5.552&& 5.558& 5.546&& 5.558& 5.546\\
1/50& 5.559&5.544&& 5.558& 5.532&& 5.557& 5.532\\
\bottomrule \end{tabular}
\end{center}
\caption{Average $\text{LogS}_{\text{tail},1}$ and $\text{LogS}_{\text{tail},2}$ for the simulation study where only the location parameter is spatially varying. Results are presented for various choices of the percentage of observed $Z_2$ realizations.} 
\label{Tab1:SimulationStudy_LogStail}
\end{table}

\begin{table}[H]
\singlespacing
\begin{center} 
\begin{tabular}{ccccccccc} \toprule 
&\multicolumn{2}{c}{Joint Asym} && \multicolumn{2}{c}{Joint Sym}&& \multicolumn{2}{c}{Ind Sym}\\
\cmidrule{2-3}\cmidrule{5-6}\cmidrule{8-9}
\makecell{Percentage\\Observed}&$\text{rMSE}_1$ & $\text{rMSE}_2$ && $\text{rMSE}_1$&$\text{rMSE}_2$&& $\text{rMSE}_1$&$\text{rMSE}_2$\\
\midrule
1/10& 26.303&27.459&& 26.301& 27.460&& 26.300& 27.485\\
1/15& 29.727&26.142&& 29.725& 26.143&& 29.725& 26.175\\
1/25& 30.711&28.905&& 30.707& 28.906&& 30.707& 28.945\\
1/35& 26.001&27.316&& 25.997& 27.311&& 25.997& 27.360\\
1/50& 27.698&28.835&& 27.696& 28.836&& 27.696& 28.873\\
\bottomrule \end{tabular}
\end{center}
\caption{Mean $\text{rMSE}_{\text{tail},1}$ and $\text{rMSE}_{\text{tail},2}$ for the simulation study where only the location parameter is spatially varying. Results are presented for various choices of the percentage of observed $Z_2$ realizations.} 
\label{Tab1:SimulationStudy_rmsetail}
\end{table}

\section{Two Alternative Proofs of Positive Semidefiniteness}\label{sec:proof_prop:RJMCMC}

\noindent\textbf{Alternative Proof 1:} 
Notice that $K^\intercal K$ is always positive semidefinite since $x^\intercal K^\intercal K x = \|Kx\|^2$ and this will be $>0$ provided $Kx \neq 0$ and therefore, this will be $>0$ for all $x \neq 0$ if $K$ is full rank.

Then, we have that $\rho \begin{bmatrix} L_1 \\ L_2 \end{bmatrix}  [L_1^\top, L_2^\top]  = \begin{bmatrix} \rho L_1L_1^\top &\rho L_1 L_2^\top \\ \rho L_2 L_1^\top &\rho L_2L_2^\top \end{bmatrix}$ will be positive definite if $[L_1^\top, L_2^\top]$ has full rank provided $\rho > 0.$ Clearly $\begin{bmatrix} (1-\rho) L_1L_1^\top &0 \\0 &(1-\rho) L_2L_2^\top \end{bmatrix} > 0$ provided each $L_1$ and $L_2$ with full rank and $\rho < 1.$ This is because every positive scalar multiple of a positive definite matrix is also positive definite. Therefore, $\begin{bmatrix} L_1L_1^\top &\rho L_1 L_2^\top \\ \rho L_2 L_1^\top &L_2L_2^\top \end{bmatrix}$ will be positive definite, being the sum of two such matrices, if all conditions are satistisfied, which reduce to $0 < \rho < 1$ and $L_1^\top, L_2^\top, [L_1^\top, L_2^\top]$ having full rank.

\bigskip

\noindent\textbf{Alternative Proof 2:} Let $\boldsymbol M := LL^T$ and look at the product $\textbf{x}^T \boldsymbol M \textbf{x}$. For ease of dealing with the blocks say that $\textbf{x} = \begin{bmatrix} x \\ y \end{bmatrix}.$ Now let, $u=L_1^Tx$ and $v=L_2^Ty$. Then we have that 

$$\textbf{x}^T \boldsymbol M \textbf{x} = u^Tu + \rho( u^Tv + v^Tu) + v^Tv.$$ 

\noindent We can rewrite this as: 
$$(1-\rho)(u^Tu + v^Tv)  + \rho(u+v)^T(u+v) \geq 0$$ 

\noindent This completes the proof.

\section{Computation Times}

Table \ref{Tab1:SimulationStudy_CompTimes} displays the average computation time for each method in seconds. Additionally, we report the computation time for one implementation of the traditional linear model of coregionalization. It can be noted that the proposed joint asymmetric model is more computationally expensive to fit than the joint and independent symmetric model counterparts; however, the difference in computation time is relatively small. Notably, each method is significantly faster than a traditional coregionalization model, which would be prohibitive to fit with 5,000 spatial locations.

\begin{table}[H]
\singlespacing
\begin{center} 
\begin{tabular}{ccccc} \toprule 
\makecell{Percentage\\Observed}&\makecell{Joint\\Asym}& \makecell{Joint\\Sym}&\makecell{Independent\\Sym (baseline)}&Full GP (baseline) \\
\midrule
1/10& 2070&1918&1971&1,436,150\\
1/15& 1873&1550&1567&591,600\\
1/25& 1305&1195&1210&234,750\\
1/35& 1436&1243&1268&191,600\\
1/50& 1213&1198&1184&179,450\\
\bottomrule \end{tabular}
\end{center}
\caption{Average computation time (seconds) for the simulation study with spatially varying location parameter. Results are presented for various choices of the percentage of observed $Z_2$ realizations.} 
\label{Tab1:SimulationStudy_CompTimes}
\end{table}

\section{Uncertainty Quantification}

The proposed methodology will quantify the uncertainty in hazard projections through the generation of an ensemble of hazard trajectories. Scenario-based uncertainty arises from unpredictable climate variations and future meteorological factors, including anthropogenic forcings. The hazard model will provide projections of tail-area behavior on both regional and local scales. Leveraging a Bayesian hierarchical framework allows for uncertainty quantification by producing an ensemble, or posterior distribution, of hazard projections, as opposed to a single estimate. The resulting output will consist of ensembles of plausible hazard projections, including return levels, return periods, and $q$-th quantile levels, among other parameters.

\section{Dependence Across Basis Functions Yields Asymmetric Cross-Covariance}

Over the past two decades, numerous studies have focused on addressing the asymmetry feature of multivariate random fields. Despite the utility of these methods, they often come with restrictive assumptions, such as stationary cross-covariance structures. One way to account for asymmetry in the cross-covariance while maintaining nonstationarity is to model dependence across basis functions. To illustrate how this yields an asymmetric cross-covariance structure, consider the cross-covariance of $\boldsymbol v(\boldsymbol s)$, which becomes

\begin{align*}
\mathsf{Cov}(v_1(\boldsymbol s),v_2(\boldsymbol t))
&=\mathsf{Cov}(\boldsymbol\Phi_1^\top(\boldsymbol s)\boldsymbol\delta_1,\boldsymbol\Phi_2^\top(\boldsymbol t)\boldsymbol\delta_2)\\
&=\boldsymbol\Phi_1^\top(\boldsymbol s)\mathsf{Cov}(\boldsymbol\delta_1,\boldsymbol\delta_2)\boldsymbol\Phi_2(\boldsymbol t)\\
&=\boldsymbol\Phi_1^\top(\boldsymbol s)\boldsymbol\Sigma_{\boldsymbol\delta_1,\boldsymbol\delta_2}\boldsymbol\Phi_2(\boldsymbol t)\\
&=\boldsymbol\Phi_1^\top(\boldsymbol s)\boldsymbol\Sigma_{\boldsymbol\delta_1,\boldsymbol\delta_2}\boldsymbol\Phi_2(\boldsymbol t)
\end{align*}

\noindent The cross-covariance of $\boldsymbol v(\boldsymbol s)$ is then

$$\boldsymbol C_v(\boldsymbol s,\boldsymbol t)=\begin{pmatrix}\boldsymbol\Phi_1^\top(\boldsymbol s)\boldsymbol\Sigma_{\boldsymbol\delta_1}\boldsymbol\Phi_1(\boldsymbol t) & \boldsymbol\Phi_1^\top(\boldsymbol s)\boldsymbol\Sigma_{\boldsymbol\delta_1,\boldsymbol\delta_2}\boldsymbol\Phi_2(\boldsymbol t) \\ \boldsymbol\Phi_2^\top(\boldsymbol s)\boldsymbol\Sigma_{\boldsymbol\delta_2,\boldsymbol\delta_1}\boldsymbol\Phi_1(\boldsymbol t) & \boldsymbol\Phi_2^\top(\boldsymbol s) \boldsymbol\Sigma_{\boldsymbol\delta_2}\boldsymbol\Phi_2(\boldsymbol t)\end{pmatrix}.$$

\noindent It follows then that the cross-covariance of $\boldsymbol w(\boldsymbol s)$ is

\begin{align*}
\boldsymbol C_w(\boldsymbol s, \boldsymbol t)
&=\boldsymbol A \boldsymbol C_v(\boldsymbol s, \boldsymbol t)\boldsymbol A^\top\\
&=\boldsymbol A\begin{pmatrix}\boldsymbol\Phi_1^\top(\boldsymbol s) \boldsymbol\Sigma_{\boldsymbol\delta_1}\boldsymbol\Phi_1(\boldsymbol t) & \boldsymbol\Phi_1^\top(\boldsymbol s)\boldsymbol\Sigma_{\boldsymbol\delta_1,\boldsymbol\delta_2}\boldsymbol\Phi_2(\boldsymbol t) \\ \boldsymbol\Phi_2^\top(\boldsymbol s)\boldsymbol\Sigma_{\boldsymbol\delta_2,\boldsymbol\delta_1}\boldsymbol\Phi_1(\boldsymbol t) & \boldsymbol\Phi_2^\top(\boldsymbol s) \boldsymbol\Sigma_{\boldsymbol\delta_2}\boldsymbol\Phi_2(\boldsymbol t)\end{pmatrix}\boldsymbol A^\top\\
&\neq\boldsymbol A\begin{pmatrix}\boldsymbol\Phi_1^\top(\boldsymbol s) \boldsymbol\Sigma_{\boldsymbol\delta_1}\boldsymbol\Phi_1(\boldsymbol t) & \boldsymbol\Phi_2^\top(\boldsymbol s)\boldsymbol\Sigma_{\boldsymbol\delta_2,\boldsymbol\delta_1}\boldsymbol\Phi_1(\boldsymbol t) \\ \boldsymbol\Phi_1^\top(\boldsymbol s)\boldsymbol\Sigma_{\boldsymbol\delta_1,\boldsymbol\delta_2}\boldsymbol\Phi_2(\boldsymbol t) & \boldsymbol\Phi_2^\top(\boldsymbol s) \boldsymbol\Sigma_{\boldsymbol\delta_2}\boldsymbol\Phi_2(\boldsymbol t)\end{pmatrix}\boldsymbol A^\top\\
&=\boldsymbol C_w^\top(\boldsymbol s, \boldsymbol t).
\end{align*}

\noindent Hence, incorporating dependence in the basis functions across processes results in an asymmetric cross-covariance structure.

\section{Construction of Joint Process for Basis Coefficients}

A challenging task is to construct a joint covariance structure $\boldsymbol\Sigma_{\boldsymbol\delta}$ for the basis coefficients that allows for asymmetric cross-covariances without violating the nonnegative definiteness of $\boldsymbol\Sigma_{\boldsymbol\delta}$.
To address this, we can generate a positive semidefinite joint covariance structure using the framework proposed by \citet{oliver2003gaussian} to explicitly model the dependence across basis functions. Consider the generative process 

$$\delta_1=\mu_1+L_1 Z_1,$$
$$\delta_2=\mu_2+L_2(\rho Z_1+\sqrt{1-\rho^2}Z_2),$$

\noindent where $L_1$ and $L_2$ are square roots of the covariance matrices, i.e.,

$$L_1 L_1^\top=C_{11}\quad\text{and}\quad L_2 L_2^\top=C_{22},$$

\noindent $\mu_1$ and $\mu_2$ are the means of $\delta_1$ and $\delta_2$, $C_{11}$ and $C_{22}$ are the covariances of $\delta_1$ and $\delta_2$, and $Z_1$ and $Z_2$ are vectors of i.i.d. standard normal random variables. It can be shown that the covariances of $\delta_1$ and $\delta_2$ are given by

$$\mathsf{Cov}(\delta_1)=L_1 L_1^\top$$
$$\mathsf{Cov}(\delta_2)=L_2 L_2^\top.$$

\noindent The cross-covariance of $\delta_1$ and $\delta_2$ is 

\begin{align*}
\mathsf{Cov}(\delta_1,\delta_2)
&=\mathbb E\left[(\delta_1-\mu_1)(\delta_2-\mu_2)^\top\right]\\
&=\mathbb E\left[(L_1 Z_1)(L_2(\rho Z_1+\sqrt{1-\rho^2}Z_2))^\top\right]\\
&=\rho L_1\mathbb E\left[Z_1 Z_1^\top\right]L_2^\top+\sqrt{1-\rho^2}L_1\mathbb E\left[Z_1 Z_2^\top\right]L_2^\top\\
&=\rho L_1 L_2^\top.
\end{align*}

\noindent It follows that if both $C_{11}=L_1 L_1^\top$ and $C_{22}=L_2 L_2^\top$ are positive definite, then so is

$$L L^\top=\begin{pmatrix}
L_1 L_1^\top & \rho L_1 L_2^\top\\
\rho L_2 L_1^\top & L_2 L_2^\top
\end{pmatrix}.$$

\noindent We can see this by noting that the matrix $L L^\top$ can be expressed as $\rho K^TK +(1-\rho) D$ with $K:=(L_1 \; L_2)$ and $D$ being block diagonal with diagonal blocks $L_iL_i^T$. Since the sum of positive definite matrices is positive definite, and since Gramians (matrices of the form $K^TK$ for any $K$) are always positive definite, and since block diagonal matrices with positive definite blocks are positive definite, and since positive definite matrices scaled by non-negative scalars are positive definite, then $L L^\top$ is positive for all $\rho\geq -1$. 

\section{Positive Semidefiniteness in Covariance Structure}

Since the joint covariance structure for the basis coefficients is positive definite, then the joint latent spatial processes for the location and scale parameters also have a positive semidefinite covariance structure, as required for model fitting. To see this, let $(\boldsymbol s_1,\ldots,\boldsymbol s_n)$ and $(\boldsymbol t_1,\ldots,\boldsymbol t_m)$ be the spatial locations associated with each of the two processes (e.g., precipitation and temperature). Let $\boldsymbol v_1=(v_1(\boldsymbol s_1),\ldots,v_1(\boldsymbol s_n))$ and $\boldsymbol v_2=(v_2(\boldsymbol t_1),\ldots,v_2(\boldsymbol t_m))$ be the latent process (i.e., location or scale processes) at these locations, with $r=n+m$ as before. The joint process for $\boldsymbol v_1$ and $\boldsymbol v_2$ can be written as

$$\begin{pmatrix}
\boldsymbol v_1\\ \boldsymbol v_2
\end{pmatrix}=\begin{pmatrix}
\boldsymbol A_{11} & \boldsymbol 0\\
\boldsymbol A_{21} & \boldsymbol A_{22}
\end{pmatrix}\begin{pmatrix}
\boldsymbol\Phi_{1} & \boldsymbol 0\\
\boldsymbol 0 & \boldsymbol\Phi_{2}
\end{pmatrix}\begin{pmatrix}
\boldsymbol\delta_{1}\\
\boldsymbol\delta_{2}
\end{pmatrix},$$

\noindent where $\boldsymbol A_{11}$, $\boldsymbol A_{21}$, and $\boldsymbol A_{22}$ are defined as before. Then the covariance of the joint process \((\boldsymbol v_1, \boldsymbol v_2)\) is given by \(\boldsymbol A\boldsymbol\Phi\boldsymbol\Sigma\boldsymbol\Phi^\top\boldsymbol A^\top\), where $\boldsymbol\Sigma=L L^\top$ is the covariance matrix of $(\boldsymbol\delta_1,\boldsymbol\delta_2)$. Since $\boldsymbol\Sigma$ is positive semidefinite, it follows that for any $\boldsymbol u\in\mathbb R^{p+q}$ that $\boldsymbol u^\top\boldsymbol\Sigma\boldsymbol u\geq 0$. 
Hence, $\boldsymbol a^\top\boldsymbol A\boldsymbol\Phi\boldsymbol\Sigma\boldsymbol\Phi^\top\boldsymbol A^\top\boldsymbol a\geq 0$ for any $\boldsymbol a\in\mathbb R^r$. Hence $\boldsymbol A\boldsymbol\Phi\boldsymbol\Sigma\boldsymbol\Phi^\top\boldsymbol A^\top$ is positive semidefinite. 

\section{Real World Example: Spatially Varying Location and Scale Parameters}

We present the results for the real-world example where both the location and scale parameters vary spatially. Figures \ref{fig:RL_precip_locscale} and \ref{fig:RL_temp_locscale} below show the predicted 10-year return levels for precipitation and temperature, respectively. The precipitation return levels are largest in southern Alabama and southern Mississippi, while western North Carolina, western South Carolina, and eastern Tennessee have the smallest predicted return levels. For temperature, the return levels are largest in northeastern Florida, with other regions exhibiting similar temperature return levels. Table \ref{Tab1:realworld_app_locscale} shows the performance results for the joint model and the independent model. The results are similar, indicating that there is not much benefit in modeling these outputs jointly.

\begin{figure}[h]
    \centering
    \begin{subfigure}[t]{0.45\textwidth}
        \centering
        \includegraphics[width=1\linewidth]{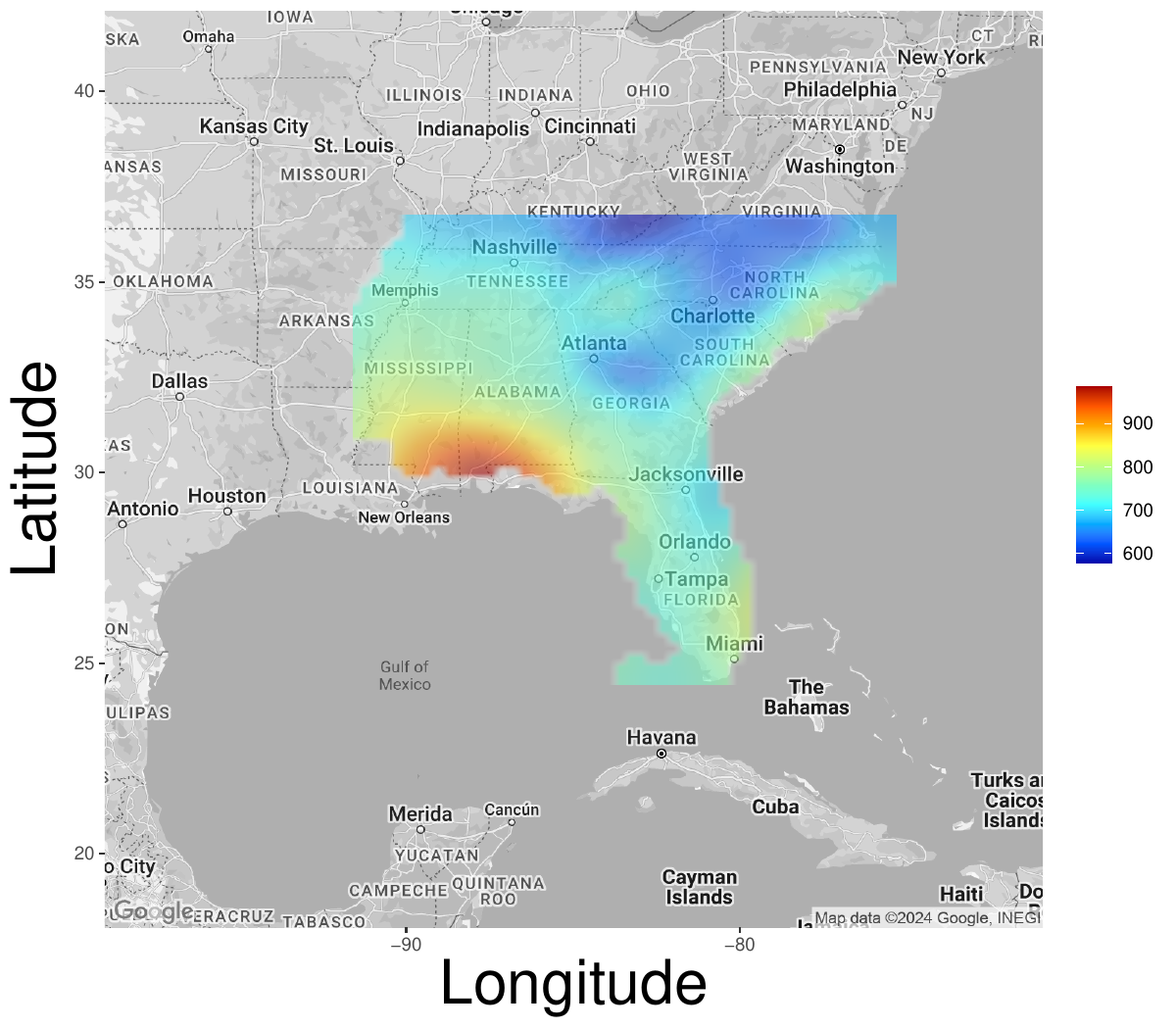} 
        \caption{Precipitation Return Levels} \label{fig:RL_precip_locscale}
    \end{subfigure}
    \hfill
    \begin{subfigure}[t]{0.45\textwidth}
        \centering
        \includegraphics[width=1\linewidth]{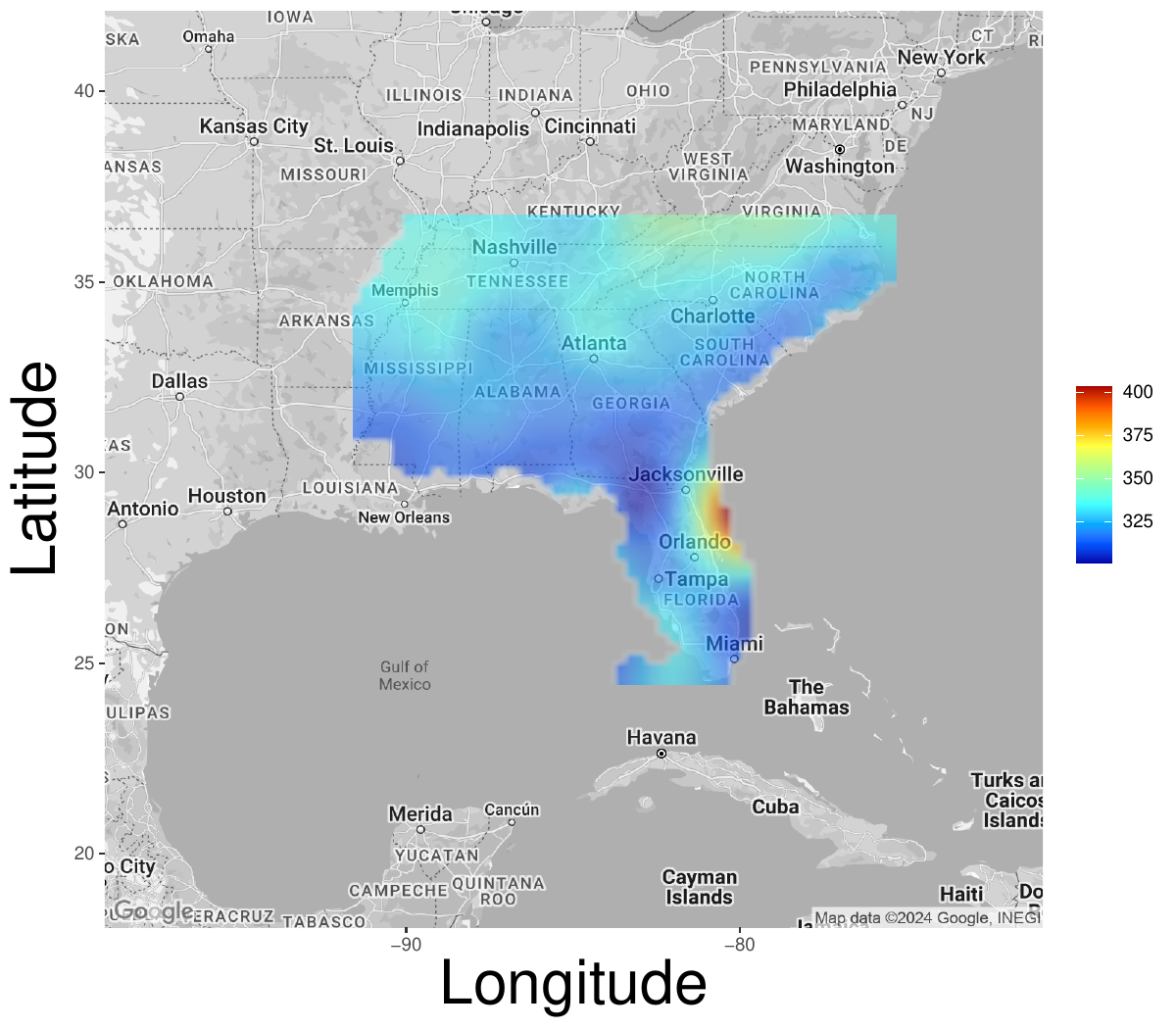} 
        \caption{Temperature Return Levels} \label{fig:RL_temp_locscale}
    \end{subfigure}
    \caption{Precipitation 10-year return levels (tenths of millimeters) (left) and temperature 10-year return levels (tenths of degrees Celsius) (right).}
\end{figure}

\begin{table}[h]
\begin{center} 
\begin{tabular}{cccccc} \toprule 
&\multicolumn{2}{c}{Joint Model} && \multicolumn{2}{c}{Independent}\\
\cmidrule{2-3}\cmidrule{5-6}
Metric&Precipitation & Temperature && Precipitation&Temperature\\
\midrule
CRPS &85.97& 22.26&&85.94&22.10\\
LogS &6.47&5.21&&6.47&5.18\\
$\text{CRPS}_{\text{tail}}$ &547.51&44.87&&547.18&45.41\\
$\text{LogS}_{\text{tail}}$ &9.26&5.95&&9.26&5.96\\
$\text{rMSE}_{\text{tail}}$ &888.39&156.61&&887.90&156.80\\
\bottomrule \end{tabular}
\end{center}
\caption{Median $\text{CRPS}$, $\text{LogS}$, $\text{CRPS}_{\text{tail}}$, $\text{LogS}_{\text{tail}}$, and $\text{rMSE}_{\text{tail}}$ for the real-world example with spatially varying location and scale parameters. Results are presented for both the joint model and independent model, and for both precipitation and temperature outputs.} 
\label{Tab1:realworld_app_locscale}
\end{table}

\section{Simulation Study Design}

 A summary table of the simulation study design is presented in Table \ref{Tab1:SimulationStudyDesign}  below.

\begin{table}[H]
\singlespacing
\begin{center} 
\begin{tabular}{cccc} \toprule 
Scenario&Percentage Observed&Cross Covariance & Spatial Variation\\
\midrule
1&1/10& Symmetric&Location Only\\
2&1/10& Symmetric&Location and Scale\\
3&1/10& Asymmetric&Location Only\\
4&1/10& Asymmetric&Location and Scale\\
5&1/15& Symmetric&Location Only\\
6&1/15& Symmetric&Location and Scale\\
7&1/15& Asymmetric&Location Only\\
8&1/15& Asymmetric&Location and Scale\\
9&1/25& Symmetric&Location Only\\
10&1/25& Symmetric&Location and Scale\\
11&1/25& Asymmetric&Location Only\\
12&1/25& Asymmetric&Location and Scale\\
13&1/35& Symmetric&Location Only\\
14&1/35& Symmetric&Location and Scale\\
15&1/35& Asymmetric&Location Only\\
16&1/35& Asymmetric&Location and Scale\\
17&1/50& Symmetric&Location Only\\
18&1/50& Symmetric&Location and Scale\\
19&1/50& Asymmetric&Location Only\\
20&1/50& Asymmetric&Location and Scale\\
\bottomrule \end{tabular}
\end{center}
\caption{Simulation Study Summary} 
\label{Tab1:SimulationStudyDesign}
\end{table}

\section{Notation Summary}

Table \ref{Tab1:Notation} below provides a summary of the dimensions and notation.

\begin{table}[H]
\singlespacing
\begin{center} 
\begin{tabular}{cccc} \toprule 
Notation& Description\\
\midrule
$r$ & Total no. of spatial locations ($r=n+m$)\\
$n$ & No. spatial locations for first process\\
$m$ & No. spatial locations for second process\\
$p$ & No. basis functions for first process ($p\ll n$)\\
$q$ & No. basis functions for second process ($q\ll m$)\\
\bottomrule \end{tabular}
\end{center}
\caption{Notation Summary} 
\label{Tab1:Notation}
\end{table}

\section{Generalized Extreme Value (GEV) Models}

Consider \(N\) independent and identically distributed random variables \(X_i\), \(i = 1, \ldots, N\), with distribution \(F\). This approach involves dividing the sample of \(N\) observations into \(m\) subsamples of \(n\) observations each (known as \(n\)-blocks) and selecting the maximum \(M_k\) from each subsample (\(k = 1, \ldots, m\)), referred to as block maxima. The extreme values of \(F\) are then identified with the sequence \(M_k\) of block maxima, and the distribution of this sequence is analyzed. The main result of EVT is that, as $m$ and $n$ grow sufficiently large, the limit distribution of (adequately rescaled) block maxima belongs to one of three different aforementioned families: the three cases just mentioned correspond, respectively, to $\xi>0$ (called the Fr\'echet case), $\xi=0$ (Gumbel) and $\xi<0$ (Weibull).

\section{Summer 2010-24 Results}

\begin{table}[h]
\vspace*{3\baselineskip}
\caption{Summer 2010-24 Results} 
\begin{center} 
\begin{tabular}{ccccccccc} \toprule 
&\multicolumn{2}{c}{Joint Model} && \multicolumn{2}{c}{Joint Sym} &&\multicolumn{2}{c}{Independent}\\
\cmidrule{2-3}\cmidrule{5-6}\cmidrule{8-9}
Metric&Precip. & Temp. && Precip. & Temp. &&Precip. & Temp.\\
\midrule
CRPS &86.47& 8.49&&86.46&8.57&&86.60&8.62\\
LogS &6.46&4.25&&6.46&4.27&&6.47&4.29\\
$\text{CRPS}_{\text{tail}}$ &556.66&24.19&&561.17&24.33&&561.30&24.85\\
$\text{LogS}_{\text{tail}}$ &9.30&5.35&&9.31&5.35&&9.30&5.38\\
$\text{rMSE}_{\text{tail}}$ &870.66&42.55&&874.83&43.08&&876.16&44.18\\
\bottomrule \end{tabular}
\end{center}
\label{Tab1:realworld_app}
\vspace*{3\baselineskip}
\end{table}

\newpage
\bibliographystyle{apalike}
\bibliography{References}

